\documentclass[english,aps,prb,superscriptaddress]{revtex4-1}
\usepackage{babel}
\usepackage{amsmath,mathtools}
\usepackage{amssymb}
\usepackage{graphicx}

\begin{document}
\title{Transport across a topoelectrical Weyl semimetal heterojunction}

\author{S M Rafi-Ul-Islam}
\selectlanguage{english}%
\affiliation{Department of Electrical and Computer Engineering, National University of Singapore, Singapore}

\author{Zhuo Bin Siu}
\email{elesiuz@nus.edu.sg}
\selectlanguage{english}%
\affiliation{Department of Electrical and Computer Engineering, National University of Singapore, Singapore}

\author{Mansoor B.A. Jalil}
\email{elembaj@nus.edu.sg}
\selectlanguage{english}%
\affiliation{Department of Electrical and Computer Engineering, National University of Singapore, Singapore}

\begin{abstract}
We propose a general method to realize and calculate the transmission in a Weyl semimetal (WSM) heterostructure by employing a periodic three-dimensional topoelectrical (TE) circuit network. By drawing the analogy between inductor-capacitor circuit lattices and quantum mechanical tight-binding (TB) models, we show that the energy flux in a TE network is analogous to the probability flux in a TB Hamiltonian. TE systems offer a key advantage in that they can be easily tuned to achieve different topological WSM phases simply by varying the capacitances and inductances. The above analogy opens the way to the study of tunneling across heterojunctions separating different types of WSMs in TE circuits, a situation which is virtually impossible to realize in physical WSM materials. We show that the energy flux transmission in a WSM heterostructure depends highly on the relative orientation of the transport direction and the $k$-space tilt direction. For the transmission from a Type I WSM source lead to a Type II WSM drain lead, all valleys transmit equally when the tilt and transmission directions are perpendicular to each other. In contrast, large inter-valley scattering is required for transmission when the tilt and transport directions are parallel to each other, leading to valley-dependent transmission. We describe a Type III WSM phase intermediate between the Type I and Type II phases. An `anti-Klein' tunneling occurs between a Type I source and Type III drain where the transmission is totally suppressed for some valleys at normal incidence. This is in direct contrast to the usual Klein tunneling in Dirac materials where normally incident flux is perfectly transmitted.  Owing to the ease of fabrication and experimental accessibility, TE circuits offer an excellent testbed to study the extraordinary transport phenomena in WSM based heterostructures.     
\end{abstract}
\maketitle

\section{Introduction} 
Studies of the topological states of matter in various platforms, such as photonic\cite{RafiRef1,RafiRef2}, mechanical\cite{RafiRef3,RafiRef4}, and ultra-cold atom systems\cite{RafiRef5,RafiRef6}, and metamaterials\cite{RafiRef7,RafiRef8} and electrical networks\cite{RafiRef9,RafiRef10,RafiRef11}, have emerged as one of the most exciting fields in contemporary condensed matter physics. Perhaps one of the most iconic features of such topological states is the existence of Weyl points in three-dimensional momentum space, at which electron and hole bands touch each other in the presence of either broken time reversal or crystal inversion symmetries. Materials with linear band dispersion in the vicinity of such exceptional points are generally classified as Weyl semimetals (WSM)\cite{RafiRef12,RafiRef13,RafiRef14}. WSMs exhibit many novel characteristics such as negative magnetoresistance\cite{RafiRef15,RafiRef16}, exotic Fermi arc surface states\cite{RafiRef14,RafiRef17,RafiRef18, SciRep7_4030}, Klein tunnelling\cite{RafiRef19,RafiRef20,RafiRef21}, quantum anomalous Hall effect\cite{RafiRef22,RafiRef23,RafiRef24} and chiral magnetic effect\cite{RafiRef25,RafiRef26}. These outstanding and fascinating properties make WSMs ideal candidate materials for next-generation nanoelectronics and quantum computing\cite{RafiRef27, JAP112_244303, APL111_063101,SciRep9_4480}. 

A new class of WSM state with anisotropic dispersion that explicitly violates Lorentz invariance was recently proposed\cite{RafiRef28,RafiRef29}. The new WSM state, named the Type II WSM phase\cite{RafiRef30,RafiRef31}, exhibits many distinctive characteristics like anisotropic magnetoresistance\cite{RafiRef32} and anisotropic chiral anomaly\cite{RafiRef33,RafiRef34}. The existence of different types of WSMs based on their dispersion tilts opens yet another avenue for exploring their potential in device applications. However, the fabrication complexity of WSM material systems and lattice structural restrictions in tuning their properties and modulating their transport behaviour present major challenges in the realization of device applications based on WSMs.
 
To overcome these limitations and explore new possibilities in WSMs, we consider a system known as a topoelectrical (TE) circuit consisting of electrical components such as inductors and capacitors to not only model different WSM states, but also to study the transport phenomena between them. TE circuits, being constructed out of basic passive electrical elements, offer the key advantage over condensed matter systems in that the system properties can be very easily tuned simply by changing the admittances of the circuit elements and the connections between them. The close correspondence between the TE network with the quantum tight binding (TB) model has motivated studies on the analogues of topological insulator states\cite{RafiRef36}, quantum spin Hall states\cite{RafiRef37}, topological corner states\cite{RafiRef9,RafiRef10} and Chern insulator states\cite{RafiRef38} based on electrical networks. Furthermore, the possible electrical detection (i.e. current, phase and impedance detection) of the Berry curvature\cite{RafiRef41}, topological edge states\cite{RafiRef10}, band structure\cite{RafiRef42} and the topological nodal states\cite{RafiRef37} in WSMs has been proposed in TE models. However, the transport between WSM phases has not yet been modelled in the corresponding TE circuits. Such transport studies in TE circuits would offer a unique platform into the topological behaviour of WSM phases. For instance, the study of the effects of the tilt strength and direction on the transport properties in a WSM heterojunction may reveal many exotic features ranging from valley-selective transmission to large inter-valley scattering. In condensed matter systems, it may not be possible to fabricate clean interfaces nor manifest the Type I and Type II WSM phases simultaneously due to the different environmental conditions (e.g. temperature, strain, pressure, etc.) needed to exhibit these states in host material systems. In contrast, TE circuits can be implemented even on simple printed circuit boards, and offer much flexibility in tuning properties which allow for sharp and clean interfaces between different WSM phases.

In this work, we establish the analogy between a TE circuit and a quantum mechanical TB Hamiltonian for a condensed matter system. To validate the experimental feasibility of our TE circuit model, we give an explanation of the transformation of an infinite lattice chain into a finite one in both one and multi-dimensional systems. We also explain in detail how to construct the TE analogues of Type I and Type II WSMs. Finally, we clarify the meaning of ``transport" and ``transmission" in the context of TE circuits. While transport carries the meaning of electron flux in condensed matter systems, we show that the analogous quantity in TE circuits is the flux of energy. This crucial analogy allows us to study the all-important transport characteristics of WSM condensed matter systems by linking them to the energy flux transmission in TE circuits. With the theoretical framework connecting TE circuits with TB models of condensed matter systems in place, we then calculate the analogue of the quantum mechanical transmission in a WSM heterojunction and reveal a key role played by the Dirac cone tilt direction in the transmission process for both types of WSMs. The transmitted energy flux across a TE analogue of a WSM heterojunction exhibits significant differences in the two cases where the transmission direction is parallel to the tilt direction, and where the transmission direction is perpendicular to the tilt direction. In the former case, the transmission shows a highly asymmetric valley profile where the energy flux transmission is significantly blocked in some valleys and only small transmission channels via inter-valley scattering are allowed. For the latter case, the transmission of energy flux shows identical profiles for all valleys. The transmission across a heterojunction separating different phases of WSM (i.e. Types I and II) has unique characteristics which can be utilized in WSM-based nanoelectronics (e.g. to realize valleytronic applications). Thus, our analysis shows that TE circuits can be designed to provide an effective testbed for studying transport properties in WSM devices before they are realized in real material systems.

To investigate the exotic transport properties at the heterojunction between different WSM phases, we consider the intermediate phase between Types I and II WSM, which we term as Type III WSM. The Type III WSM state shows a flat admittance band dispersion along its tilt direction. As a result, transmission across a heterojunction into a Type III WSM drain exhibits a unique transport behaviour where total internal reflection occurs for some states at normal incidence to the interface, which we term as ``anti-Klein'' tunneling. This is in contrast to the maximum transmission at normal incidence for the case of transmission into a Type II WSM drain lead, which corresponds to the normal Klein tunneling. This pronounced difference in transport phenomena between the two different drain types is due to the fact that propagating states exist for only one of the two pseudospin branches in a Type III WSM, while in the case of Type II WSM, propagating modes exist for both of the pseudospin branches. The pseudospin mismatch between the incident source modes in some valleys and the drain modes results in the complete suppression of tunneling in the anti-Klein effect.

\section{Analogy between a TE circuit and a quantum mechanical system}

\subsection{Hamiltonian analogue}
We explain how to establish the analogy between an inductance-capacitance (LC) circuit, and the quantum mechanical TB Hamiltonian. In complex circuit theory, the complex current $I$ flowing through a LC component is given by $I = Y\delta V$ where $Y$ is the admittance of the component, and $\delta V$ is the potential difference across the component. The admittance of a capacitor with capacitance $C$ for a harmonic AC current of angular frequency $\omega$ is $Y_C = i\omega C$, while that of an inductor with inductance $L$ is $Y_L = -\frac{i}{\omega L} $. From the perspective of admittance, an inductor therefore behaves like a capacitor with a negative capacitance, with the induction corresponding to a negative capacitance $-|C|$ given by $|L|^{-1} = (\omega^2) |C|$. In the rest of this paper we shall, for simplicity refer to capacitors and capacitance exclusively with the understanding that a negative capacitance actually refers to an inductance. 

\begin{figure}[ht!]
\centering
\includegraphics[scale=0.6]{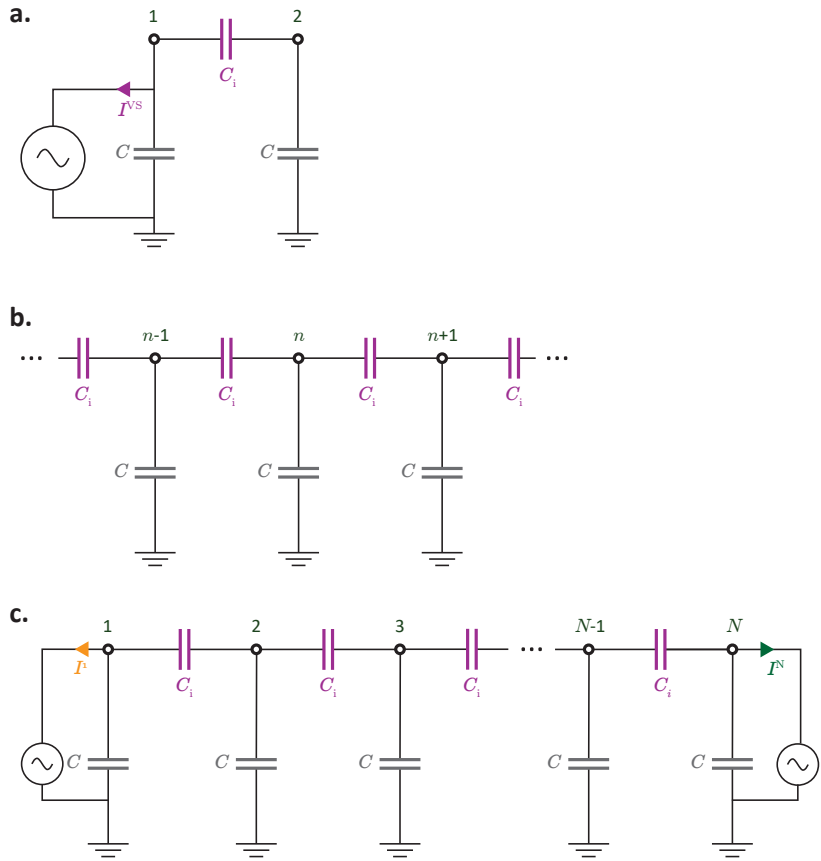}
\caption{  a. Schematic of a two-node circuit consisting of two voltage nodes connected to the ground via identical capacitances $C$, and connected to each other via a coupling capacitance $C_{\mathrm{i}}$. A voltage source is connected to node 1 with the current flowing through the voltage source denoted as $I^{\mathrm{VS}}$. b. Schematic of an infinite chain of voltage nodes coupled to each other with coupling capacitances $C_{\mathrm{i}}$ and connected to the ground via common capacitances $C$. c. Schematic of a finite chain of voltage nodes equivalent to the infinite chain in panel b, consisting of the chain truncated to $N$ nodes and voltage supplies attached to both ends of the finite chain.  } 
\label{gFig1}
\end{figure}

To set the stage, let us consider a very simple system illustrated in panel a. of Fig. \ref{gFig1}, consisting of two circuit nodes numbered 1 and 2 connected by identical capacitors $C$ to the ground. The two nodes are connected by a capacitor $C_{\mathrm{i}}$. (The $\mathrm{i}$ in the subscript is not to be interpreted as a dummy index.) An ideal voltage source sets the voltage bias at node 1 to $V_1$ with respect to the ground. We denote the current flowing through the voltage supply as $I^{\mathrm{VS}}$. 

In general, the total current flowing into the $n$th voltage node in a circuit composed purely of capacitors (and inductors), $I_n$, is given by 
\[
	I_n = (i\omega) \sum_{a} C_{an} (V_n - V_a)
\]
where the sum runs over all the nodes which node $n$ is connected to via the coupling capacitances $C_{an}$. By Kirchoff's Current Law (KCL), the net current flowing into any voltage node is 0. We therefore have 
\begin{equation}
	0 = \sum_a C_{an} (V_n - V_a) - \frac{I^{\mathrm{E}}_n}{i\omega} \label{KCL}
\end{equation} 
where $I^{\mathrm{E}}_n$ accounts for any additional current flows out of the $n$th node besides those due to capacitors connected to thoe node. In the circuit in Fig. \ref{gFig1}a, $I^{\mathrm{E}}_1 = I^{\mathrm{VS}}$, $I^{\mathrm{E}}_2 = 0$. 
Applying Eq. \ref{KCL} to the two nodes in Fig. \ref{gFig1} panel a., we have 
\begin{equation}
	C \begin{pmatrix} V_1 \\ V_2 \end{pmatrix} = C_{\mathrm{i}} \begin{pmatrix} -1 & 1 \\ 1 & -1 \end{pmatrix} \begin{pmatrix} V_1 \\ V_2 \end{pmatrix} + \frac{i}{\omega} \begin{pmatrix} I^{\mathrm{VS}} & 0 \\ 0 & 0 \end{pmatrix}. \label{Ham10} 
\end{equation}
Eq. \ref{Ham10} looks almost like an eigenvalue equation for the matrix $C_{\mathrm{i}} (-\mathbf{I}_2 + \sigma_x)$ with eigenvalues $C$ and eigenvector $(V_1, V_2)^\mathrm{T}$ if not for the second term containing the current flowing through the voltage supply. ($\mathbf{I}_2$ is the two-by-two identity matrix and $\sigma_x$ is the $x$ Pauli matrix.)  We can ask the question of what values $C$ should take for a given value of $C_{\mathrm{i}}$, so that the current flowing through the voltage supply is zero. This question is answered by setting $I^{\mathrm{VS}} = 0$ in Eq. \ref{Ham10}, so that Eq. \ref{Ham10} takes the form of
\begin{equation}
	C \mathbf{v} = C_{\mathrm{i}} (-\mathbf{I}_2 + \sigma_x) \mathbf{v} \label{Ham11}
\end{equation}
where bold uppercase symbols denote matrices and bold lowercase symbols vectors. Here $\mathbf{v}$ is the vector of the voltages at the nodes. Eq. \ref{Ham11} can readily diagonalized to give the eigenvalues $C = C_{\mathrm{i}}(1 \pm 1)$. Notice that once the $C$ capacitances are set to $C_{\mathrm{i}}(1 \pm 1)$ and the voltage bias supplied by the voltage supply $V_1$ set to an arbitrary value, Kirchoff's Laws automatically guarantee that the voltage at node $2$ will be $\pm V_1$. 

Eq. \ref{Ham11} is the TE analogue of a quantum-mechanical TB Hamiltonian consisting of only two lattice sites with on-site energy $-C_{\mathrm{i}}$ and hopping integral $C_{\mathrm{i}}$. More generally, TE analogues to quantum TB Hamiltonians can be extended to any number of lattice sites. In these TE analogues, the voltage nodes connected to the ground via the capacitance $C$ play the role of TB lattice points while the capacitive (and inductive) couplings between lattice sites play the role of TB hopping integrals. Hence, the voltage profile across the nodes in a TE circuit is analogous to the spatial variation of the wavefunction in a TB system while $C$ stands in for the eigenenergy.  For a finite one-dimensional chain with only nearest-neighbour couplings, the chain can be populated with a desired eigenmode by attaching a voltage supply at one of the voltage nodes. Generalizing Eq. \ref{Ham11}, we write 
\begin{equation}
	C \mathbf{v} = \mathbf{H} \mathbf{v} \label{Ham1}
\end{equation}
where $\mathbf{H}$ is the TE analogue of the TB Hamiltonian obtained from writing down the KCL at every node, moving the terms containing the capacitance $C$ to the left of the equal sign and setting the current flowing through the attached voltage supply to zero. We want the current through the voltage supply to be zero by design, because otherwise a finite current flow will contribute an imaginary on-site potential energy term into $\mathbf{H}$, an example of which is the $-I^{\mathrm{VS}}/i\omega$ term in Eq. \ref{Ham10}. The imaginary on-site potential energy term breaks the Hermitricity of $\mathbf{H}$. Imposing the current through the voltage supply to be zero ensures that the voltage supply does not introduce any additional terms into $\mathbf{H}$, and can thus be attached to any of the voltage nodes without affecting the form of $\mathbf{H}$. The eigenvalues of $\mathbf{H}$ hence correspond to the allowed values of $C$ in which the current flowing through the voltage supply is indeed zero. We stress that although no current flows through the voltage supply when $C$ is set to one of the eigenvalues, the physical scenario is not equivalent to simply removing the voltage supply and leaving the circuit open. The AC voltage supply imposes a finite $\frac{\mathrm{d}V}{\mathrm{d}t}$ at the node it is attached to. The finite temporal rate of change of the potential differences across the capacitors in the circuit leads to the charging or discharging of these capacitors and the flow of electrical currents through them even though the current through the voltage source is zero. 

Let us briefly summarize what we have done so far. We saw that the TE analogue Eq. \ref{Ham1} to the Schrodinger's equation $E|\psi\rangle = H|\psi\rangle$ can be established by considering the KCL at the voltage nodes of a LC circuit. The voltage profile vector $\mathbf{v}$ is the TE analogue of the wavefunction while $C$, the common grounding capacitance for all the nodes, plays the role of the eigenenergy.  In finite TE circuits analogous to finite TB systems, the eigenvalues of $C$ correspond to the values of common grounding capacitances for which no current flows through a voltage supply attached to any node. (The voltage supply serves to populate the voltage nodes with finite voltages.) The fact that no current flows through the voltage supply for an eigenvalue of $C$ allows us to attach the voltage supply to any node without modifying $\mathbf{H}$ and the voltage profile of the eigenmodes. 

\subsection{Infinite one-dimensional chains} 
Building towards our goal to study heterojunctions between semi-infinite leads, let us now move on from finite circuits and consider the infinite one-dimensional chain shown in Fig. \ref{gFig1}b. Each node is connected to its immediate  neighbours to the left and right by a coupling capacitance $C_{\mathrm{i}}$, and connected to the ground via the common grounding capacitance $C$. Writing down the KCL for the $n$th node, we have
\begin{equation}
	C V_n = C_{\mathrm{i}} ( V_{n+1} + V_{n-1} - 2 V_n). \label{kclInfChain} 
\end{equation}
This is a recursive relation in $V_n$. Substituting the ansatz that $V_n = v_0 \exp(i k n)$ into Eq. \ref{kclInfChain}, we have $C = 2 C_{\mathrm{i}} (\cos(k) - 1)$ which gives $k = \kappa \equiv \arccos(\frac{C}{2C_{\mathrm{i}}}+1)$. For a given value of $C_{\mathrm{i}}$, modifying the value of $C$ allows us to modify the spatial wavelength of the voltage profile $2\pi/\kappa$. Notice that both $V_n = v_0 \exp(i \kappa n)$ and $V_n = v_0 \exp(-i \kappa n)$ satisfy Eq. \ref{kclInfChain}. The voltage profile of an eigenmode of the infinite chain thus has the general form of 
\begin{equation}
	V_n = \alpha_+ \exp(i \kappa n) + \alpha_- \exp(-i \kappa n) \label{kclInfChainSup}
\end{equation}
where $\alpha_\pm$ are the weightages of the $\exp(\pm i \kappa n)$ eigenmodes. This is the TE analogue of a one-dimensional free electron gas system with $C$ playing the role of the eigenenergy. 

Unlike in a finite chain where a specific eigenmode can be populated simply by setting $C$ to the eigenvalue corresponding to the desired eigenmode and connecting a voltage supply to any one of the nodes, it turns out that the voltage profile in an infinite chain would not be uniquely specified by attaching a finite number of voltage biases to the nodes. (Details in the appendix.) The ability to controllably populate desired linear superpositions of eigenstates is an important requirement in setting the direction of flux flowing along a TE heterojunction -- we want to populate the eigenmodes  such that the TE analogue of probability flux flows only from the source to the drain and not vice-versa. (We will discuss the TE analogue to probability flux in the next section.)   Moreover, infinitely long chains cannot be  constructed in actual experiments. For practical purposes, it is necessary to represent an infinite or semi-infinite circuit by an equivalent circuit comprising a finite number of nodes. In the following,  we construct a chain with a \textit{finite} number of nodes in which the voltage profile in the interior has an identical form to that in an infinite chain as given by Eq. \ref{kclInfChainSup}. This is, as we shall later see, sufficient to model the transmission from a semi-infinite long source lead to a semi-infinite long drain lead in a heterojunction system. 

Fig. \ref{gFig1}c shows the finite-length chain which can be used to model the homogenous infinite-length chain in Fig. \ref{gFig1}b.  The finite-length chain is essentially the latter truncated to a finite number of nodes, $N$, where we now attach voltage supplies to each of the two ends. We number the nodes from 1 to $N$, and allow finite current flows through the voltage supplies. We denote the currents flowing through the voltage supply at node 1 ($N$) as $I^1$ ($I^N$). Consider nodes 2 to $N-1$. Each of these nodes is coupled to its left and right neighbours by $C_{\mathrm{i}}$, and to the ground via the common grounding capacitance $C$. The KCLs for these nodes take the same form as the KCLs for the infinite chain Eq. \ref{kclInfChain} for $2 \leq n \leq N-1$. The voltage profile in these nodes therefore also takes the same form as Eq. \ref{kclInfChainSup} for the infinite chain. The finite chain models the infinite chain in the sense that the voltages in the interior nodes in the former are governed by the same Hamiltonian $\mathbf{H}$ as the latter and have the same voltage profiles as the infinite chain eigenmodes. The required voltage profile in the finite chain can be obtained by setting the voltage biases at the two ends at appropriate values as will be explained below.

Let us now examine how the weights of the $\exp(\pm i \kappa n)$ modes inside the chain, $\alpha_\pm$, are related to the voltage biases $V^1$ and $V^N$ set by the voltage supplies at the two ends. At node 1, the KCL reads 
\begin{eqnarray}
	C V_1 + \frac{I^1}{i \omega} &=& C_{\mathrm{i}} (V_2 - V_1) \label{homoV1a} \\
	&=& C_{\mathrm{i}} ( \alpha_+ \exp(2 i \kappa) + \alpha_- \exp(-2 i \kappa) - V_1)  \label{homoV1b} 
\end{eqnarray}
where in going from Eq. \ref{homoV1a} to Eq. \ref{homoV1b}, we made use of the fact that $V_2$ is described by Eq. \ref{kclInfChainSup}.
Similarly, at node $N$ the KCL reads 
\begin{equation}
	C V_N + \frac{I^N}{i\omega} = C_{\mathrm{i}} (\alpha_+ \exp((N-1)i\kappa) + \alpha_-\exp(-i(N-1)\kappa) - V_N). \label{homoVN}
\end{equation} 
At node 2, the KCL reads 
\begin{eqnarray}
	&& C V_2 = C_{\mathrm{i}} (V_1 + V_3 - 2V_2) \nonumber \\
	&\Rightarrow& (C+2C_\mathrm{i})(s_+\exp(2i\kappa)+s_-\exp(-2i\kappa))  - C_{\mathrm{i}} (V_1 + s_+\exp(3i\kappa) + s_-\exp(-3i\kappa)) = 0 \label{homoV2}
\end{eqnarray}
where again  $V_2$ and $V_3$ take the form as described by Eq. \ref{kclInfChainSup}.
In a similar vein, we have at node $N-1$,
\begin{equation}
	(C+2C_\mathrm{i})(s_+\exp(i(N+1)\kappa)+s_-\exp(-i(N+1)\kappa))  - C_{\mathrm{i}} (V_N + s_+\exp(i(N-2)\kappa) + s_-\exp(-i(N-2)\kappa)) = 0.  \label{homoVNn1}
\end{equation}

Eqs. \ref{homoV1b} to \ref{homoVNn1} constitute a system of four equations relating the six variables $V_{(1/N)}$, $I^{(1/N)}$, and $\alpha_\pm$ to one another. These equations may be interpreted in two complementary ways. In the first, we take $V_1$ and $V_N$ as given values and solve for $\alpha_\pm$ and $I^{(1/N)}$. This corresponds to setting the voltage biases at the two ends of the chain to the given values, and finding out what linear superposition of the $\exp( \pm i \kappa n)$ modes result inside the chain. Alternatively, we can take $\alpha_\pm$ to be given, and solve for $V_{(1/N)}$ and $I^{(1/N)}$. This corresponds to finding out what voltage biases need to be set at the ends of the chain in order to achieve the desired linear superposition of the $\exp( \pm i\kappa n)$ modes inside the chain.  

\begin{figure}[ht!]
\centering
\includegraphics[scale=0.6]{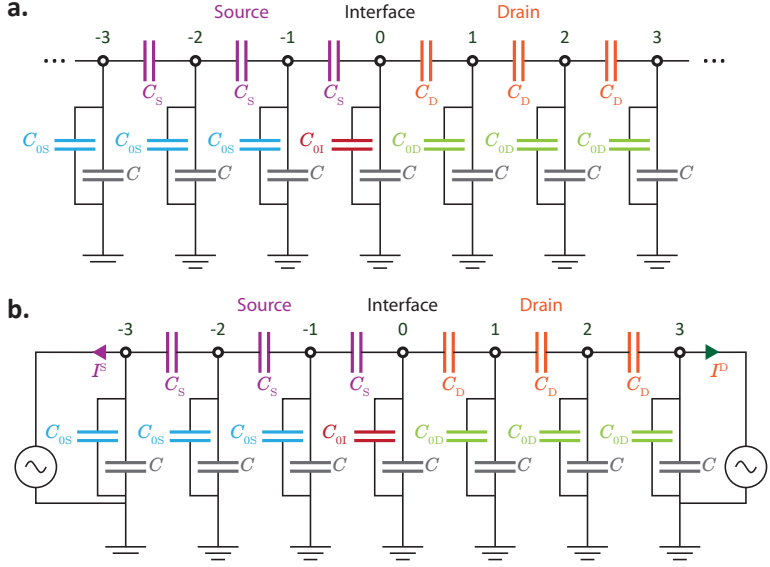}
\caption{  a. Schematic of a heterojunction between a semi-infinite source lead consisting of nodes coupled to their left and right neighbours by a capacitance $C_{\mathrm{S}}$ and to the ground via the common grounding capacitance $C$ and the `on-site energy' $C_{0\mathrm{S}}$, and a semi-infinite drain lead consisting of nodes coupled to their left and right neighbours by a capacitance $C_D$ and to the ground via the common grounding capacitance $C$ and `on-site energy' $C_{0\mathrm{D}}$. b. Schematic of a finite length chain to model the transmission through the infinite chain consisting with the source and drain leads truncated at the third node away from the interface, and voltage supplies attached to both ends of the system.   } 
\label{gFig1B}
\end{figure}

Note that a key difference between the finite length chains in panels a. and c. of Fig. \ref{gFig1} is that we have restricted the current flow through the voltage supply to zero in the former. This is equivalent to setting a hard-wall boundary condition at the ends of the finite chain. The circuit in panel a. is hence the TE analogue of a discretized infinite potential well where the eigenspectrum consists of a discrete set of eigenvalues. In contrast, allowing finite currents to flow through the voltage supplies at the ends of the chain in panel c. is equivalent to setting open boundary conditions at those ends. With reference to the Non-Equilibrium Green's Function formalism, the imaginary potentials at nodes 1 and $N$ resulting from the current flows through the voltage supplies can be taken to be the imaginary parts of the lead self-energies\cite{A2T} at the two ends of a central barrier region attached to semi-infinite leads . The imaginary potentials allow a finite-sized chain to model an infinite-sized one by folding the effects of the portions of the infinite-dimensional Hamiltonian excluded in the former into non-Hermitian terms. The $C$ eigenvalues therefore fall into a continuum. A corollary result is that a semi-infinite chain can be modeled by attaching a voltage source to only one end of a finite chain. This allows the modeling of a TB heterojunction consisting of a semi-infinite source lead connected to a semi-infinite long drain lead, such as the one shown in Fig. \ref{gFig1B}a.

\subsection{One-dimensional heterojunction}
The fact that an infinite chain can be modeled by a finite one allows us to model the transmission from a semi-infinite source lead to a different semi-infinite drain lead across a heterojunction using a finite number of nodes. Panel a. of Fig. \ref{gFig1B} shows one such heterojunction circuit with semi-infinite leads, and panel b., the finite-length circuit to model the infinite heterojunction system of panel a. 

The source lead of the heterojunction consists of nodes connected to their left and right neighbours by the coupling capacitance $C_{\mathrm{S}}$, and to the ground via the common grounding capacitance $C$ and an additional capacitance $C_{0\mathrm{S}}$ connected in parallel to $C$. $C_{0\mathrm{S}}$ plays the role of the TE analogue to a TB on-site potential energy term, as can be seen from the KCL for the $n$th source node,
\begin{equation}
	C V_n = C_{\mathrm{S}} (V_{n+1} + V_{n-1} - 2 V_n) - C_{0\mathrm{S}} V_n. \label{kclSrc}
\end{equation}
This is the TE analogue to the TB Hamiltonian $H_{\mathrm{S}} = \sum_n  (|\phi_n\rangle \langle \phi_{n+1}| + \mathrm{h.c.})C_{\mathrm{S}} - |\phi_n\rangle ( C +  2C_{\mathrm{S}}+C_{\mathrm{0S}}) \langle \phi_n|$. In an actual circuit one may, of course, combine $C$ and $C_0$ into a single physical capacitor. We have depicted them as separate capacitors to emphasize their respective roles where $C$ plays the TE analogue to the eigenenergy while $-C_{0\mathrm{S}}$ is the TE analogue to an on-site potential.  

The general form of an eigenmode in the source lead is thus
\begin{equation}
	V_n = s_+ \exp(i \kappa_{\mathrm{S}} n) + s_- \exp(-i\kappa_{\mathrm{S}} n ) \label{SrcVn}
\end{equation}
where $\kappa_{\mathrm{S}} \equiv \arccos( \frac{C + C_{0\mathrm{S}}}{2C_{\mathrm{S}}} + 1)$, and $s_\pm$ is the weightage of the $\exp(\pm i \kappa_{\mathrm{S}} n)$ mode. 

The drain lead modeled consists of nodes connected to their left and right neighbours by the coupling capacitance $C_{\mathrm{D}}$ and to the ground via the common grounding capacitance $C$ and the `on-site potential' capacitance $C_{0\mathrm{D}}$. The KCL at the $n$th drain node is 
\begin{equation}
	C V_n = C_{\mathrm{D}} ( V_{n+1} + V_{n-1} - 2 V_n ) -  C_{0\mathrm{D}} V_n. \label{kclDrn}
\end{equation}

The general solution of Eq. \ref{kclDrn} for $V_n$ in the drain lead is thus 
\begin{equation}
	V_n = d_+ \exp(i \kappa_{\mathrm{D}} n) + d_- \exp(-i\kappa_{\mathrm{D}} n) \label{DrnVn} 
\end{equation}
where $\kappa_{\mathrm{D}} \equiv \arccos (\frac{C+C_\mathrm{0D}}{2C_{\mathrm{D}}} + 1)$ and $d_\pm$ is the weightage of the $\exp(\pm i \kappa_{\mathrm{D}} n)$ mode.

Similar to Fig. \ref{gFig1}b, we model the infinite chain by truncating it to a finite number of nodes -- three on either side of the interface in this case -- and attaching voltage supplies to both ends of the chain in the finite-length model. In the circuit in Fig. \ref{gFig1B}b, nodes -2 and -1 obey the KCL Eq. \ref{kclSrc}. The voltage profile in the two nodes thus takes the form of Eq. \ref{SrcVn}. Similarly, nodes 1 and 2 obey the KCL Eq. \ref{kclDrn} and have a voltage profile given by Eq. \ref{DrnVn}. In analogy to what we did in Eqs. \ref{homoV1b} and \ref{homoVN}, the KCL at the terminal nodes -3 and 3 give
\begin{equation}
	C V_{-3} + \frac{I^{\mathrm{S}}}{i\omega} = C_{\mathrm{S}} ( s_+\exp(-2i\kappa_{\mathrm{S}}) + s_-\exp(2i\kappa_{\mathrm{S}}) -V_{-3}) - C_{0\mathrm{S}} V_{-3} \label{hetVn3}
\end{equation}
and
\begin{equation}
	C V_{3} + \frac{I^{\mathrm{D}}}{i\omega} = C_{\mathrm{D}} (d_+\exp(2i\kappa_{\mathrm{D}}) + d_- \exp(-2i\kappa_{\mathrm{D}}) - V_3 ) - C_{0\mathrm{D}} V_3 \label{hetV3}
\end{equation}
respectively. 

In a similar manner to Eqs. \ref{homoV2} and \ref{homoVNn1}, the KCLs at nodes -2 and 2 in the  source and drain yield
\begin{equation}
	(C+C_{0\mathrm{S}} + 2C_\mathrm{S})(s_+\exp(-2i\kappa_{\mathrm{S}})+s_-\exp(2i\kappa_{\mathrm{S}}))  - C_{\mathrm{i}} (V_{-3} + s_+\exp(-i\kappa_{\mathrm{S}}) + s_-\exp(-i\kappa_{\mathrm{S}})) = 0 \label{hetVn2}
\end{equation}
and
\begin{equation}
	(C+C_{0\mathrm{D}} + 2C_\mathrm{D})(d_+\exp(2i\kappa_{\mathrm{D}})+d_-\exp(-2i\kappa_{\mathrm{D}}))  - C_{\mathrm{i}} (V_{3} + d_+\exp(i\kappa_{\mathrm{D}}) + d_-\exp(i\kappa_{\mathrm{D}})) = 0 \label{hetV2} 
\end{equation}
respectively. 

Finally, the KCL at the interface between the source and drain at node 0 reads
\begin{equation}
	(C +C_{0\mathrm{I}})V_0 = C_{\mathrm{S}} (V_{-1} - V_0) + C_{\mathrm{D}} (V_1 - V_0). \label{kclV0}
\end{equation}
This differs from both Eqs. \ref{kclSrc} and \ref{kclDrn}. Therefore $V_0$ is given by neither Eqs. \ref{SrcVn} nor \ref{DrnVn}, and has to be solved for explicitly. Expanding $V_{\pm 1}$ using Eqs. \ref{SrcVn} and \ref{DrnVn}, we have

\begin{equation}
	( C+C_{0\mathrm{I}} + C_{\mathrm{S}} + C_{\mathrm{D}} )V_0 - C_{\mathrm{S}}(s_+\exp(-i\kappa_{\mathrm{S}}) + s_-\exp(i\kappa_{\mathrm{S}})) - C_{\mathrm{D}}(d_+\exp(i\kappa_{\mathrm{D}}) + d_-\exp(-i\kappa_{\mathrm{D}})) = 0 \label{hetV0}.
\end{equation} 	

On the source side of the interface, the KCL at node -1 gives
\begin{eqnarray}
	&& (C + C_{0\mathrm{S}})V_{-1} = C_{\mathrm{S}} (V_{-2} + V_0 - 2 V_{-1}) \nonumber \\
	&\Rightarrow& (C + C_{0\mathrm{S}} +  2C_{\mathrm{S}}) (s_+ \exp(-i\kappa_{\mathrm{S}}) + s_-\exp(i\kappa_{\mathrm{S}})) - C_{\mathrm{S}} (s_+ \exp(-2i\kappa_{\mathrm{S}}) + s_-\exp(2i\kappa_{\mathrm{S}}) + V_0) = 0 \label{hetVn1}.
\end{eqnarray}
Similarly, the KCL at node 1 gives
\begin{equation}
	(C +C_{0\mathrm{D}} +  2C_{\mathrm{D}}) (d_+ \exp(i\kappa_{\mathrm{D}}) + d_-\exp(i\kappa_{\mathrm{D}})) - C_{\mathrm{D}} (d_+ \exp(2i\kappa_{\mathrm{D}}) + d_-\exp(-2i\kappa_{\mathrm{D}}) + V_0) = 0 \label{hetV1}.
\end{equation} 

Eqs. \ref{hetVn3} to \ref{hetV1} together constitute a system of seven equations relating the nine variables $V_{-3 / 3}$, $I^{\mathrm{S}/\mathrm{D}}$, $s_\pm$, $d_\pm$ and $V_0$ to one another. In solving for the transmission of an incident mode from the source lead to the drain lead through the interface, we explicitly set the weightage of the source mode propagating towards the interface to a finite value which we take as 1 for convenience, and that of the drain mode propagating towards the interface to zero. This corresponds to fixing $s_+ = 1$ and $d_- = 0$. There are then seven remaining unknowns which can be solved for with the seven equations. In particular, the solved value of $s_-$ represents the reflection coefficient for propagating modes incident from the source side, $d_+$ the transmission coefficient, and $V_{\pm 3}$ the voltage biases that need to be set at the ends of the chain in order to fix $s_+$ at 1 and $d_-$ at 0. ( Setting $V_{\pm 3}$ at arbitrary voltages differing from the solution of Eq. \ref{hetVn3} to \ref{hetV1} will result in deviations of $d_-$ from zero, corresponding to the injection of flux from the drain lead to the source lead. This will reduce the net amount of flux flowing from the source lead to the drain lead.  )

We briefly explain why Eqs. \ref{hetVn3} to \ref{hetV1} applied to the various nodes in Fig. \ref{gFig1B}b would model the transmission from a truly semi-infinite source lead to a semi-infinite drain lead in Fig. \ref{gFig1B}a. If the leads have been truly semi-infinite, the transmission would have been solved by using exactly the same equations as the three equations Eq. \ref{hetV0}, \ref{hetVn1} and \ref{hetV1} with $s_+$ set to 1 and $d_-$ set to zero. The three equations provide the correct number of equations to solve for the reflection coefficient $s_-$, transmission coefficient $d_+$ and interface voltage $V_0$. On top of having the same KCL equations as the infinite-sized system at the interface (Eq. \ref{hetV0}) and the two nodes on either side of the interface (Eqs. \ref{hetVn1} and \ref{hetV1}), we also need the nodes at $n=\pm2$ in the finite model to have the same voltages as the corresponding nodes in the actual infinite-sized system. This is because the voltage at node $n=2$ ($n=-2$) also appears in Eq. \ref{hetV1} (Eq. \ref{hetVn1}) alongside the voltages at nodes 0 and 1 (-1).  Consider, for example, the drain side. We ensured the voltage at the $n=2$ nodes match in the finite and infinite systems when we derived Eq. \ref{hetV1} by expressing the voltage at the $n=2$ node in the finite-length model in the form of $V_n = d_+ \exp(i\kappa_{\mathrm{D}}n) + d_-\exp(-i\kappa_{D}n)$, as it would if the system had been infinite in size. Node 2 serves as an intermediate buffer between the interface and node 3 to which the voltage supply is attached. The voltage at node 3 does not necessarily follow the form of $V_n = d_+ \exp(i\kappa_{\mathrm{D}}) + d_-\exp(-i\kappa_{D})$ since there are additional terms in the KCL there resulting from the current flowing through the voltage supply. 

Apart from Eqs. \ref{hetV0} to \ref{hetV1} which have the same forms in both the finite-length model and the  infinite-length system, we have additionally applied Eqs. \ref{hetVn3} to \ref{hetV2} to the former in order to relate $s_+$ and $d_-$ to the terminal voltages $V_{\pm 3}$, since $s_+$ and $d_-$s cannot be directly manipulated in experiments. The terminal voltages $V_{\pm 3}$ can be directly adjusted in order to make $s_+$ and $d_-$ attain the desired values of $1$ and $0$ respectively. 

\subsection{Multi-dimensional TE circuits}
We have so far discussed one-dimensional chains. Let us now consider modeling the transmission in higher-dimensional heterojunctions, since we will subsequently be modeling WSMs, which are three-dimensional systems.

\begin{figure}[ht!]
\centering
\includegraphics[scale=0.6]{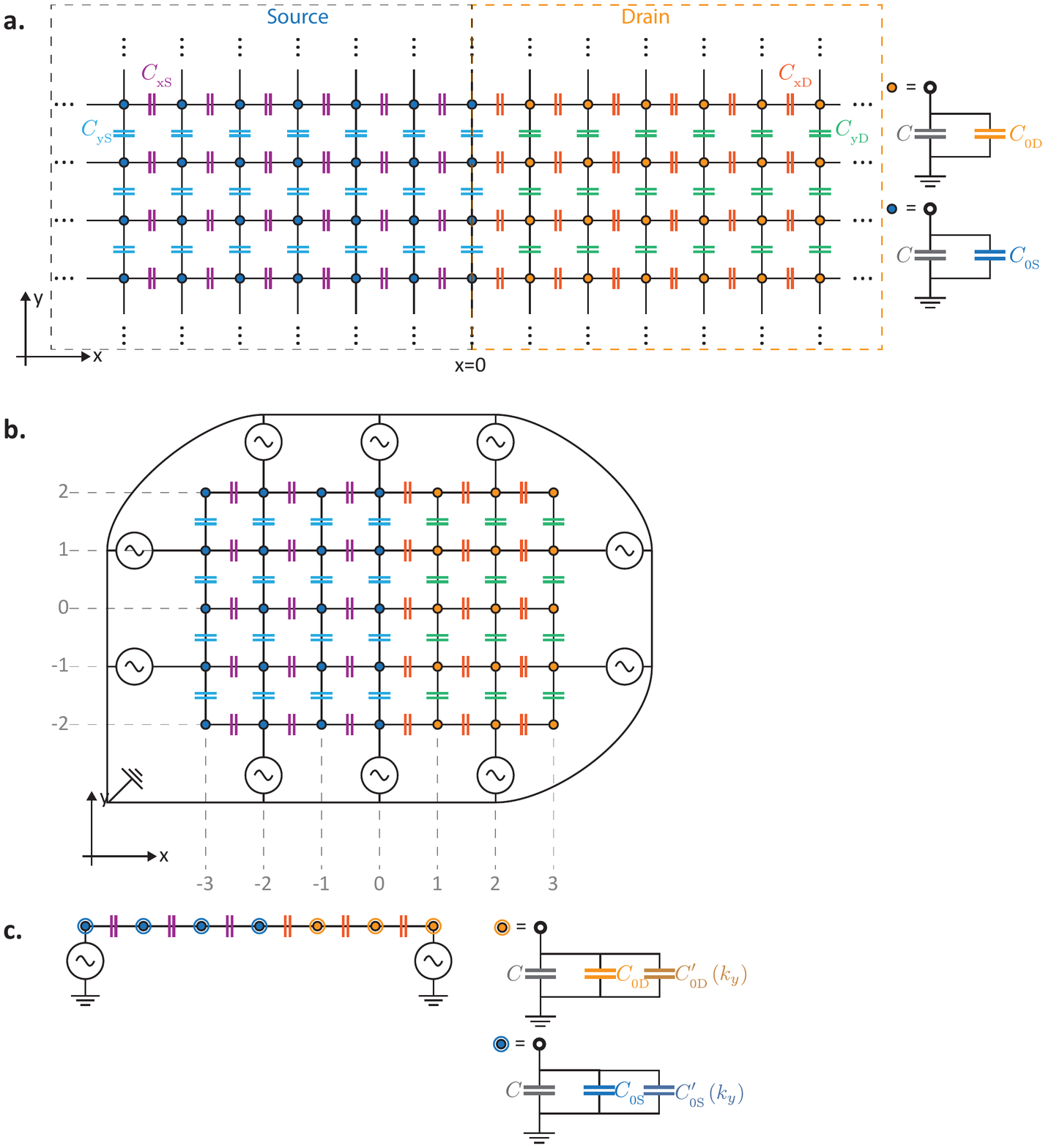}
\caption{  a. A two-dimensional heterojunction between a semi-infinite source lead consisting of nodes coupled to their left and right neighbours by a capacitance $C_{\mathrm{xS}}$, upper and lower neighbours by a capacitance $C_{\mathrm{yS}}$,  and to the ground via the common grounding capacitance $C$ and the `on-site energy' $C_{\mathrm{0S}}$, and a semi-infinite drain lead consisting of nodes coupled to their left and right neighbours by a capacitance $C_{\mathrm{xD}}$, upper and lower neighbours by a capacitance $C_{\mathrm{yD}}$ and to the ground via the common grounding capacitance $C$ and `on-site energy' $C_{\mathrm{0D}}$. b. Schematic of a finite-sized TE circuit to model the transmission through the infinite heterojunction with the source and drain leads truncated at three nodes away from the interface perpendicular to the interface and at five nodes parallel to the interface, and voltage supplies attached to every other node along the outer edge of the circuit. c. The finite two-dimensional circuit of b. can be further reduced to a finite one-dimensional TE circuit shown systematically here. The circuit consists of one row of the infinite chain with an additional $k_y$ dependent capacitance attached in parallel to the common grounding capacitance at each node.    } 
\label{gFig1C}
\end{figure}	

Fig. \ref{gFig1C}a  shows an exemplary two-dimensional heterojunction formed from the interface between a semi-infinite source lead extending to $x\rightarrow -\infty$ and a semi-infinite drain lead extending to $x\rightarrow +\infty$. The source and drain leads also extend to $\pm \infty$ along the transverse $y$ direction parallel to the interface. The source lead (drain lead) consists of nodes coupled to their left and right neighbours by the capacitance $C_{\mathrm{xS}}$ ($C_{\mathrm{xD}}$) and to their upper and lower neighbours by $C_{\mathrm{yS}}$ ($C_{\mathrm{yD}}$). Each node is connected to the ground via the common grounding capacitance $C$ and an `on-site potential' capacitance $C_{\mathrm{0(S/D)}}$.  

Let us consider the KCL at a source node.  Labeling the nodes by their $(x,y)$ coordinates so that the voltage at the node at $(x,y)$ is denoted as $V_{x,y}$, we have
\begin{equation}
	(C + C_{\mathrm{0S}}) V_{x,y} = C_{\mathrm{xS}} (V_{x+1,y} + V_{x-1,y} - 2 V_{x,y}) + C_{\mathrm{yS}} (V_{x,y+1}+V_{x,y-1} - 2 V_{x,y}). \label{kclVxyS}  
\end{equation} 
The solution to Eqn. \ref{kclVxyS} takes the form of $V_{x,y} \simeq \exp(i (k_x x + k_y y))$ where $k_x$ and $k_y$ satisfy
\begin{equation}
	(C + C_{\mathrm{0S}} + 2C_{\mathrm{xS}} + 2C_{\mathrm{yS}}) = 2 (C_{\mathrm{xS}} \cos(k_x) + C_{\mathrm{yS}} \cos(k_y)). \label{kxy} 
\end{equation}
In transmission problems, the conserved momentum parallel to the interface, $k_y$, is assumed to be a given quantity. Denoting the two solutions of $k_x$ to Eq. \ref{kxy} for a given $C$ and $k_y$  as $\pm \kappa_{\mathrm{xS}}$, a general solution to Eq. \ref{kclVxyS} within the source with the given value of $k_y$ takes the form of 
\begin{equation}
	V_{x,y} = \exp(i k_y y) ( s_+ \exp(i \kappa_{\mathrm{xS}} x) + s_- \exp(-i \kappa_{\mathrm{xS}} x)) \label{VxyS}
\end{equation} 
where $s_\pm$ are the coefficients of the $\exp(i (k_y y \pm \kappa_{\mathrm{xS}} x))$ eigenmodes.
Analogously we write a general solution to Eq. \ref{kclVxyS} within the drain as \ 
\begin{equation}
	V_{x,y} = \exp(i k_y y) ( d_+ \exp(i \kappa_{\mathrm{xD}}x ) + d_- \exp(-i \kappa_{\mathrm{xD}} x)) \label{VxyD}
\end{equation} 
where $\kappa_{\mathrm{xD}}$ is the drain analogue to $\kappa_{\mathrm{xS}}$ calculated from Eq. \ref{kclVxyS} with $C_{\mathrm{0S}}$, $C_{\mathrm{xS}}$ and $C_{\mathrm{yS}}$ replaced by $C_{\mathrm{0D}}$, $C_{\mathrm{xD}}$ and $C_{\mathrm{yD}}$ respectively. 

Similar to the discussion following Fig. \ref{gFig1B}b, we want to set up the TE circuit so that the voltage profile in the source has $s_+ = 1$ in Eq. \ref{VxyS} and that in the drain has $d_- = 0$ in Eq. \ref{VxyD} in order to model the physcial scenario of modes incident on the interface from the source side. As before, we would also want to model the two-dimensional heterojunction using a finite number of nodes, since it is not experimentally practical to construct semi-infinite long leads. 

Fig. \ref{gFig1C}b shows a finite circuit which models the heterojunction. Similar to the circuit in Fig. \ref{gFig1B}b, it comprises of the infinite-sized circuit truncated down to three nodes on both sides of the interface along the longitudinal transport direction, to which the voltage supplies are attached to at both ends. The reason for having seven nodes along the transport direction for the two-dimensional case here is identical to that for the one-dimensional case in Fig. \ref{gFig1B}b: The KCLs at the nodes along the interface at $x=0$, and the nodes to the immediate left and right of the interface at $x=\pm 1$ are identical for the infinite heterojunction in Fig. \ref{gFig1C}a and its finite-sized model in panel Fig. \ref{gFig1C}b. Solving the KCLs for the finite-sized model will therefore yield the same solutions as solving the infinite system provided that the voltages at the $x=\pm 2$ nodes have the same values they would have in the actual infinite-sized system. The nodes at $x=\pm 2$ also follow the same KCLs (Eq. \ref{kclVxyS} and its drain analogue) as they would in the infinite system, and serve as intermediates between the nodes adjacent to the interface and the nodes at $x=\pm 3$, to which the voltage supplies are attached and which would therefore have different KCLs from their corresponding nodes in the infinite system.  

The same reasoning also underlies why the circuit in Fig. \ref{gFig1C}b is truncated at five nodes along the $y$ direction. The nodes at $y=0$ and $\pm 1$ have the same KCLs and voltage profiles as the infinite system in order for the finite system to model the infinite sized one. The nodes at $\pm 1$ mediate between the inner nodes and the perimeter nodes at $y=\pm 2$ to which voltage supplies are attached. 

Let us consider the requisite number of voltage sources for the finite two-dimensional TE circuit of Fig. \ref{gFig1C}b to represent the infinite two-dimensional TE heterojunction circuit of Fig. \ref{gFig1C}a with $s_+ = 1$ and $d_- = 0$ (i.e. forward transmission from source to drain). There are 20 nodes along the perimeter of the system at $\big(x \in (-3,...,3), y= \pm 2 \big)$ and $\big (x= \pm 3, y \in (-2,...,2) \big)$, for which 20 equations can be written down based on the KCL at each of these nodes. Note that not all of these perimeter nodes need to be affixed to voltage sources. For those nodes which have attached voltage supplies, the KCLs will resemble Eqs. \ref{hetVn3} and \ref{hetV3} and contain terms corresponding to the currents flowing through the voltage supplies. The voltages at the other perimeter nodes without any attached voltage supplies would have to be solved for explicitly since these nodes are coupled to only one node along the direction perpendicular to the edge, unlike the nodes in the interior of the circuit. There are another 12 nodes one layer in at $\big (x \in (-2,...,2), y = \pm 1 \big)$ and $\big(x = \pm 2, y \in (-1,..,1)\big)$ acting as intermediates between the outer nodes with attached voltage supplies and the inner-most nodes which have the same KCLs, voltage profiles and neighbours as the infinite system. One KCL resembling Eqs. \ref{hetVn2} and \ref{hetV2} can be written for each of these nodes. Finally there are the three inner-most nodes at $\big( x = (-1,0,), y=0 \big)$ for which three KCLs resembling Eqs. \ref{hetV0} to \ref{hetV1} can be written. There are therefore a total of 20 + 12 + 3 = 35 KCL equations which can be written, one for each node. 

Let us now look at the number of unknowns to be solved for. The 20 voltages at the perimeter nodes along the edges, and the voltages at the three nodes at $\big(x=0, y\in(-1,0,1)\big)$ along the interface not already included amongst the perimeter nodes, have to be solved for, giving 20+3 = 23 unknowns. $s_-$ and $d_+$, the reflection and transmission coefficients for modes incident on the interface from the left, constitute another two unknowns. There is a remainder of 35 - 25 = 10 unknowns required in order for the number of unknowns needed to match the number of equations. These 10 unknowns are provided by attaching 10 voltage supplies along the outer edges of the circuit, where the current flowing through each voltage supply provides for an additional unknown. For symmetry, voltage supplies are attached to every other node along the outer edge of the circuit. (In general, any 10 of the perimeter nodes can be chosen.) The circuit in Fig. \ref{gFig1C}b therefore has a matching number of unknowns and equations which can be uniquely solved for the voltage biases to be set at each voltage supply so that $s_+ = 1, d_- = 0$. 

The schematic in Fig. \ref{gFig1C}b, while faithful to the modeled infinite heterojunction in Fig. \ref{gFig1C}a by being comprised of a finite section of the heterojunction, may not be very convenient to construct experimentally in practice. A more straightforward, but perhaps less instructive, implementation is to note that as far as the transmission is concerned the effects of the spatial dimensions transverse to the transmission direction can be incorporated as an `on-site energy' term. This then allows the finite circuit in panel b. to be further reduced to a one-dimensional circuit shown in panel c. comprising of two voltage sources and a single row of capacitors.  To see how the dimension parallel to the interface (i.e. parallel to the $y$ direction) can be reduced into a one-dimensional array of on-site potentials, consider, for example, the KCL at node $(x=-1,y=0)$ in the infinite circuit: 
\begin{eqnarray}
	&& C V_{x=-1,y=0} \nonumber \\
	&=& C_{\mathrm{xS}} ( V_{x=-2,y =0} + V_{x=0,y=0} - 2 V_{x=-1,y=0}) + C_{\mathrm{yS}} (V_{x=-1,y=1} + V_{x=-1,y=-1} - 2 V_{x=-1,y=0}) - C_{\mathrm{0S}} V_{x=-1,y=0} \label{eff2D1} \\
	&=& C_{\mathrm{xS}} ( V_{x=-2,y =0} + V_{x=0,y=0} - 2 V_{x=0,y=0}) - V_{x=-1,y=0} (C_{\mathrm{0S}} - [2 C_{\mathrm{yS}} (\cos(k_y)-1) ]) \label{eff2D2}
\end{eqnarray}
In going from Eq. \ref{eff2D1} to Eq. \ref{eff2D2} we made use of the fact that Eq. \ref{VxyS} implies that $V_{x=-1,y=\pm 1} = \exp( \pm i k_y) V_{x=-1,y=0}$. The terms in the square bracket in Eq. \ref{eff2D2} due to the coupling along the $y$ direction are mathematically identical to the terms that would result from attaching  an additional on-site capacitance $C'_{0\mathrm{S}}(k_y) = -2 C_\mathrm{yS}(\cos(k_y)-1)$ to the node $C_{x=-1,y=0}$, had the node been along a one-dimensional chain. Extending this analogy to the other nodes along $y=0$, we see that higher-dimensional TE heterojunctions can be reduced to a one-dimensional chain of panel c. via appropriate choices of $C_{0(\mathrm{S,I,D})}$ to incorporate the effects of the dimensions parallel to the interface. 
 
\subsection{Current analogue}
In the previous subsections, we have set up the framework of a TE circuit made of capacitive elements and driven appropriately by voltage sources to model the injected charge ``transport"  from the source to the drain of a TB heterojunction. In this section, we discuss the physical meaning of the transport in our TE circuit and how it can be linked to the electron transport in a TB Hamiltonian. In a TB Hamiltonian, the hopping of the charge carriers between lattice sites gives rise to a conserved probability flux of charge carriers. Here, we derive the analogous quantity in a TE circuit, and explain its physical meaning. We will calculate the TE analogue of the probability flux in WSM heterojunctions in a later section.   
 
Let us recall the derivation of the probability flux in a TB system. We begin with the familiar continuum one-dimensional case from elementary quantum mechanics. Consider an eigenstate $|\psi^{\mathrm{C}}\rangle$ of a time-independent Hamiltonian $H^{\mathrm{C}}$ satisfying $i\partial_t |\psi^{\mathrm{C}}\rangle = H^{\mathrm{C}}|\psi^{\mathrm{C}}\rangle = |\psi^{\mathrm{C}}\rangle E$ where $E$ is the eigenenergy of the state $|\psi^{\mathrm{C}}\rangle$. The superscript $\mathrm{C}$ serves as a reminder that these are continuum quantities. $|\psi^{\mathrm{C}}\rangle$ is related to its wavefunction $\psi^{\mathrm{C}}(x)$ via 
\begin{eqnarray}
	|\psi^{\mathrm{C}}\rangle &=& \int\mathrm{d}x\ |x\rangle \langle x|\psi^{\mathrm{C}}\rangle \nonumber \\
	&=& \int\mathrm{d}x\ |x\rangle \psi^{\mathrm{C}}(x) \label{psix} 
\end{eqnarray}
where the wavefunction $\psi^{\mathrm{C}}(x) \equiv \langle x|\psi^{\mathrm{C}}\rangle$ is formally defined as the inner product of the state vector $|\psi^{\mathrm{C}}\rangle$ with the state localized at the spatial position $x$, $|x\rangle$. The rate of change of particle density at $x$ is given by 
\begin{eqnarray}
	\partial_t |\psi^{\mathrm{C}}(x)|^2 &=&   \langle \partial_t \psi^{\mathrm{C}}|x\rangle \psi^{\mathrm{C}}(x) + \psi^{\mathrm{C}}(x)^*\langle x|\partial_t \psi^{\mathrm{C}} \rangle \nonumber \\
	&=& \frac{1}{i} \big(\psi^{\mathrm{C}}(x)^*\langle x| (H^{\mathrm{C}}|\psi^{\mathrm{C}}\rangle) - (\langle \psi^{\mathrm{C}}|H^{\mathrm{C}})|x\rangle\psi^{\mathrm{C}}(x) \big)  \label{dtrho}.
\end{eqnarray}

Let us now consider the somewhat less familiar case of a \textit{discretised} one-dimensional TB system. We consider a homogenous one-dimensional TB chain with a time-independent Hamiltonian
\begin{equation}
	H = c \sum_n ( |\phi_{n+1}\rangle \langle \phi_n| + \mathrm{h.c.}) \label{disQMH} 
\end{equation}
where $c$ is the hopping coupling between the lattice sites which we take to be real, and $|\phi_n\rangle$ is the discrete analogue of $|x\rangle$, representing the state localized at the $n$th lattice site. An arbitrary TB state $|\psi\rangle$ may then be written, analogously to Eq. \ref{psix}, as
\begin{equation}
	|\psi\rangle = \sum_n |\phi_n\rangle\langle\phi_n|\psi\rangle = \sum_n |\phi_n\rangle \psi_n
\end{equation} 
where $\psi_n \equiv \langle \phi_n|\psi\rangle$ is the discrete analogue of the wavefunction $\psi(x)$ indexed by the discrete lattice site index $n$ rather than the continuum $x$. 
Analogous to the derivation of Eq. \ref{dtrho}, the rate of change of particle density at the lattice site $n$, $\rho_n \equiv |\psi_n|^2$, is given by 
\begin{equation}
	\partial_t \rho_n = \frac{1}{i}  ( \psi_n^* \langle \phi_n|H|\psi\rangle -\langle\psi|H|\phi_n\rangle \psi_n ) \label{QMdtp1}.
\end{equation}

Let $|\psi\rangle$ be an eigenstate of $H$ satisfying $H|\psi\rangle = |\psi\rangle E$. From the definition of $H$ in Eq. \ref{disQMH}, the $\langle \phi_n|H|\psi\rangle$ term that appears in Eq. \ref{QMdtp1} is equal to
\begin{eqnarray}
	\langle \phi_n|H|\psi\rangle &=& \langle \phi_n| ( c \sum_m ( |\phi_{m}\rangle (\psi_{m+1} + \psi_{m-1}) ) \nonumber \\
	&=& c (\psi_{n+1} + \psi_{n-1}) \label{phinHpsi}.
\end{eqnarray}
Substituting Eq. \ref{phinHpsi} into Eq. \ref{QMdtp1} gives
\begin{eqnarray}
	\partial_t |\psi_n|^2 &=&  \frac{1}{i} c ( \psi_n^*(\psi_{n+1} + \psi_{n-1}) - (\psi_{n+1}^* + \psi_n)\psi_n ) \nonumber \\
	&=& -(   \mathrm{Im} (2 c \psi_{n+1}^*\psi_{n} )  - \mathrm{Im} ( 2 c \psi_{n}^*\psi_{n-1} ) ) \label{dtRhoDivJ}.
\end{eqnarray} 

Interpreting Eq. \ref{dtRhoDivJ} as the discrete one-dimensional version of the continuity equation $\partial_t \rho_n = -\nabla\cdot\vec{j}_n$, $2c\, \mathrm{Im}(\psi_{n+1}^*\psi_n)$ carries the physical interpretation of the probability flux flowing from the $n$th lattice site to its neighbour on the right.  On the other hand, $|\psi\rangle$ satisfying $H|\psi\rangle = |\psi\rangle E$ implies that the right hand side of the equal sign in Eq. \ref{QMdtp1} reduces to $(-E + E) |\psi_n|^2 = 0$. Thus, $2c \, \mathrm{Im}(\psi_{n+1}^*\psi_n) =  2c\, \mathrm{Im}(\psi_{n}^*\psi_{n-1})$ for all $n$. This implies that the probability flux at $n$, $j_n$,
\begin{equation}
	j_n = \mathrm{Im} (2 c \psi_{n+1}^*\psi_n) \label{TBjn}, 
\end{equation}
is a conserved quantity that has the same value throughout the entire length of the chain.   

We now proceed to find a conserved quantity analogous to the probability flux in a TE circuit. The form of Eq. \ref{QMdtp1} comprising of a $\langle \psi|H|\phi_n\rangle \psi_n $ and its complex conjugate suggests the consideration of the TE analogue to $\langle \psi|H|\phi_n\rangle \psi_n$. Let $\mathbf{v}$ be a TE eigenmode of $\mathbf{H}$ satisfying Eq. \ref{Ham1}, $ \mathbf{v}C = \mathbf{H}\mathbf{v}$. Eq. \ref{Ham1} is analogous to the Schroedinger equation $|\psi\rangle C = H|\psi\rangle$ with $\mathbf{v}$ being the analogue of $|\psi\rangle$. The TE analogue to the TB wavefunction at site $n$, $\psi_n$, is thus $V_n$, the voltage at node $n$. $\langle \psi|H|\phi_n\rangle$ is equal to $\langle \phi_n|H|\psi\rangle^*$, the complex conjugate of $n$th element of the matrix $H|\psi\rangle$. The TE analogue of the latter is $(\mathbf{H}\mathbf{v})_n$, the $n$th element of the matrix product of $\mathbf{H}$ and $\mathbf{v}$.   Taken together, the TE analogue of $\langle \psi|H|\phi_n\rangle\psi_n \rightarrow (\mathbf{H}\mathbf{v})_n^*V_n$. 

$(\mathbf{H}\mathbf{v})_n^*V_n$ by itself is not a physically meaningful quantity. We however note that an eigenmode $\mathbf{v}$ satisfies $(\mathbf{H}\mathbf{v})_n = C\mathbf{v}_n$ and that $\frac{C\mathbf{v}_n}{i\omega}$ is the current from node $n$ through $C$ to the ground. $\frac{1}{i\omega}(\mathbf{H}\mathbf{v})_n^*V_n$ is thus the product of a voltage and a current, and represents the complex power flowing from node $V_n$ to the ground. In analogy to Eq. \ref{QMdtp1}, we therefore consider half the sum of $-(\frac{\mathbf{H}\mathbf{v}}{i\omega})_n^*V_n$ and its complex conjugate. This represents the real part of the complex power from the ground to node $n$. The power is itself the rate of change of energy $E$. We thus have 
\begin{equation}
	\partial_t E = \frac{1}{2i\omega} (  \mathbf{v}_n^*(\mathbf{H}\mathbf{v})_n - (\mathbf{H}\mathbf{v})_n^*\mathbf{v}_n ). \label{TEdtrho1} 
\end{equation}

We expand the right hand side of Eq. \ref{TEdtrho1} using $(\mathbf{H}\mathbf{v})_n = CV_n$. Let us consider a generic case of $CV_n = C_{n;n+1} V_{n+1} + C_{n;n-1} V_{n-1} + C_0 V_n$ for a one-dimensional chain with only nearest-neighbour coupling. This corresponds physically to a voltage node coupled by a capacitance $C_{n;n-1}$ to its left neighbour, $C_{n;n+1}$ to its right neighbour and $C_0$ to the ground. Eq. \ref{kclV0} is one example where $C_{n;n-1} \neq C_{n;n+1}$. Substituting $(\mathbf{H}\mathbf{v})_n = C_{n;n+1} V_{n+1} + C_{n;n-1} V_{n-1} + C_0 V_n$ into Eq. \ref{TEdtrho1}, we obtain 
\begin{equation}
	\partial_t E = -\frac{1}{\omega} ( \mathrm{Im} (C_{n+1;n} V_{n+1}^* V_n - C_{n;n-1} V_n^* V_{n-1} ). \label{TEdtrho2}  
\end{equation}
Interpreting the right hand side of Eq. \ref{TEdtrho2} as the discretised version of the one-dimensional divergence, Eq. \ref{TEdtrho2} also takes the form of the continuity equation $\partial_t \rho = -\nabla\cdot\vec{j}$ where in the case of Eq. \ref{TEdtrho2}, $\rho = E$. We thus identify $\frac{1}{\omega} \mathrm{Im} ( C_{n+1;n} V_{n+1}^* V_n )$ as the energy flux flowing from $V_n$ to $V_{n+1}$. The energy flux is a conserved quantity for an eigenmode. This can be seen by considering Eq. \ref{TEdtrho1}: Substituting $\mathbf{H}\mathbf{v}_n = CV_n$ into the equation, the right hand side reduces to $C |V_n|^2 - C |V_n|^2 = 0$. The left hand side of Eq. \ref{TEdtrho2} is thus zero for all values of $n$, and the energy flux has the same constant value along the entire chain. The energy flux at node $n$, $j_{\mathrm{E};n}$
\begin{equation}
	j_{\mathrm{E};n} = \frac{1}{\omega} \mathrm{Im}\ (C_{n+1;n} V_{n+1}^*V_n) \label{TBjEn}
\end{equation}
is therefore the TE analogue of the TB probability flux Eq. \ref{TBjn} and can be calculated at any node $n$ to obtain its conserved value along the chain. 
We note one important difference between the TE system and a TB one. Whereas in the latter the wavefunction at a lattice site is not a directly measurable quantity by itself (only expectation values of an observable operator acting on the wavefunction are measurable), the voltage at each node in a TE system can be directly measured by attaching a voltmeter to the node. The energy flux can therefore be calculated via Eq. \ref{TBjEn} from the measured voltages across any two adjacent nodes. 

\section{WSMs in TE circuits} 
Having constructed the TE circuit equivalent of a TB model of a heterojunction and established the equivalence between the energy flux in the TE circuit with the electron flux in the TB model, we are now in a position to study TE circuit analogues of WSMs.

\begin{figure}[ht!]
\centering
\includegraphics[scale=0.6]{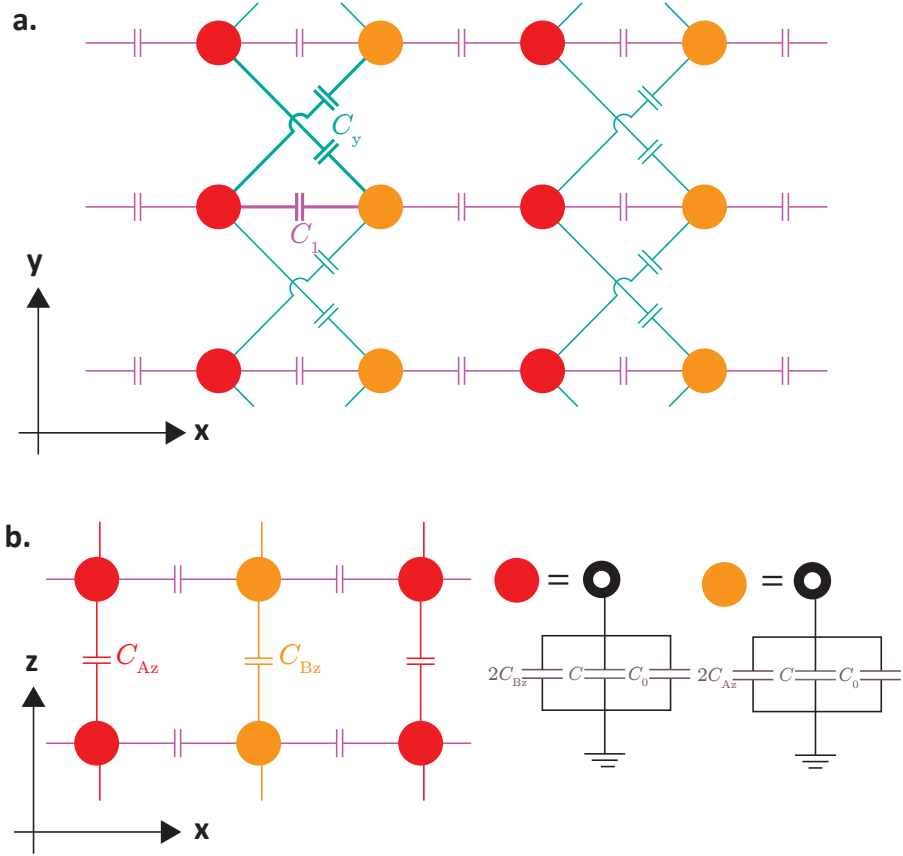}
\caption{ a. Schematic of the $xy$ plane of a TE circuit hosting the WSM state.  b. Schematic of the $xz$ plane of the same TE circuit.  The red and orange filled circles denote the A and B sublattice sites; each filled circle consists of a voltage node connected to the ground with an `on-site potential' capacitance $C_0$ and the common grounding capacitance $C$. Additionally, each A and B type node is grounded by a capacitor $2 C_{Bz}$and $2 C_{Az}$ respectively.}
\label{gFig2}
\end{figure}

Appropriately constructed TE circuits can host the analogue of the WSM phase in which the $C$-bands disperse linearly with $\vec{k}$ along all three spatial dimensions in the vicinity of the band-touching Weyl points. (Note that the $C$-bands of a TE circuit are the analogue of energy dispersion of a TB Hamiltonian.) We consider the circuit depicted in Fig. \ref{gFig2} consisting of a regular array of repeating units along the $x$, $y$ and $z$ directions. Each unit consists of a A sublattice site and B sublattice site, with each site comprising of a voltage node connected to the ground via the common grounding capacitance $C$ and an `on-site potential' capacitance $C_0$. Additionally, each A and B type node is grounded by a capacitor with capacitance $2 C_{Bz}$ and $2 C_{Az}$ respectively.  The system has reflection symmetry along the $z$ and $y$ directions.

Denoting the A/B sublattice degree of freedom as the Pauli matrices $\sigma_i$s, the TE Hamiltonian for the circuit in Fig. \ref{gFig2} reads 
\begin{eqnarray}
	H(\vec{k}) &=& ( (C_{Az}+C_{Bz})\cos(k_z)  )\mathbf{I}_\sigma - ( C_1(1+\cos(k_x) )+ C_y(2\cos(k_y)) )\sigma_x \nonumber \\
	&& - C_1\sin(k_x)\sigma_y - (C_{Az}-C_{Bz})\cos(k_z)\sigma_z  \nonumber \\ 
	&& +(2C_1+2C_y+2(C_{Az}+C_{Bz})+C_0)\mathbf{I}_\sigma. 	\label{WSMham}  
\end{eqnarray}

For convenience, we eliminate the terms in the last line of Eq. \ref{WSMham} by setting $C_0 = -(2C_1+2C_y+2(C_{Az}+C_{Bz}))$ so that the Weyl points where the hole and particle bands touch each other occur at $C=0$. 

The Weyl points of Eq. \ref{WSMham} correspond to the points in $k$-space where the coefficients of $\sigma_x$, $\sigma_y$ and $\sigma_z$ simultaneously vanish. Denoting these points as $\vec{k}_0 = (k_{x0},k_{y0},k_{z0})$, we have  
\begin{eqnarray}
	C_1\sin(k_{x0})&=&0, \label{Wpkx} \\
	(C_{Az}-C_{Bz})\cos(k_{z0}) &=& 0, \label{Wpkz} \\
	C_1 (1 + \cos(k_{x0}) ) + C_y (2\cos(k_{y0})) &=& 0. \label{Wpky}
\end{eqnarray}
Eq. \ref{Wpkx} implies that $k_{x0}  = \pi$. ( $k_{x0} = -\pi$ is also a solution of Eq. \ref{Wpkx}, but $k_{x0}=\pm \pi$ are equivalent if we set the Brillouin zone boundaries at $\pm \pi$. )  Eq. \ref{Wpkz} implies that $k_{z0} = \pm \pi/2$.  Substituting the values of $k_{x0}$ and $k_{z0}$ into Eq. \ref{Wpky} gives $k_{y0} = \pm \pi/2$. There are therefore 4 inequivalent Weyl nodes located at $(k_{x0},k_{y0},k_{z0})=\pi(1, \eta_y /2, \eta_z /2)$ where $\eta_y$ and $\eta_z$  assume values of $\pm 1$.  

Considering now the linear expansion of Eq. \ref{WSMham} for a small displacement $\delta\vec{k} = (\delta k_x, \delta k_z, \delta k_y)$ around $\vec{k}_0 = \frac{\pi}{2}(2, \eta_y,\eta_z)$, we have 
\begin{equation}
	H(\vec{k}_0 + \delta\vec{k}) = -\begin{pmatrix} 2C_1\sigma_x \\ 2\eta_y C_y\sigma_x  \\  \eta_z ( (C_{Az} + C_{Bz})\mathbf{I}_\sigma + (C_{Az}-C_{Bz}) \sigma_z ) \end{pmatrix} \cdot \begin{pmatrix} \delta k_x \\ \delta k_y \\ \delta k_z \end{pmatrix}. \label{linHam}
\end{equation}

The last term, i.e., the coefficient of  $\delta k_z$ contains a $(C_{Az}-C_{Bz})$ term which is proportional to the Fermi velocity along the $k_z$ direction, and a $(C_{Az} + C_{Bz})\mathbf{I}_\sigma$ term which imparts a tilt to the Dirac cone along the $k_z$ direction. Our TE model will host Type I Weyl nodes as long as the coefficient of $\sigma_z$ dominates over the coefficient of $\mathbf{I}_\sigma$. Conversely, a Type II WSM phase emerges when $|C_{Az} + C_{Bz} | > | C_{Az}-C_{Bz}|$. Moreover, a new topological WSM phase we call the Type III phase emerges at the transition point between the Type I and Type II phases when $| C_{Az} + C_{Bz} | = | C_{Az}-C_{Bz}|$.  Summarising, the conditions for the three WSM phases are 
\begin{eqnarray}
	&& |C_{Az} + C_{Bz} | < | C_{Az}-C_{Bz}|,\ \text{Type I}  \label{T1Cond}  \\
	&& |C_{Az} + C_{Bz} | > | C_{Az}-C_{Bz}|,\ \text{Type II}   \label{T2Cond} \\
	&& |C_{Az} + C_{Bz} | = | C_{Az}-C_{Bz}|, \ \text{Type III} \label{T3Cond}
\end{eqnarray}

\begin{figure}[ht!]
\centering
\includegraphics[scale=0.6]{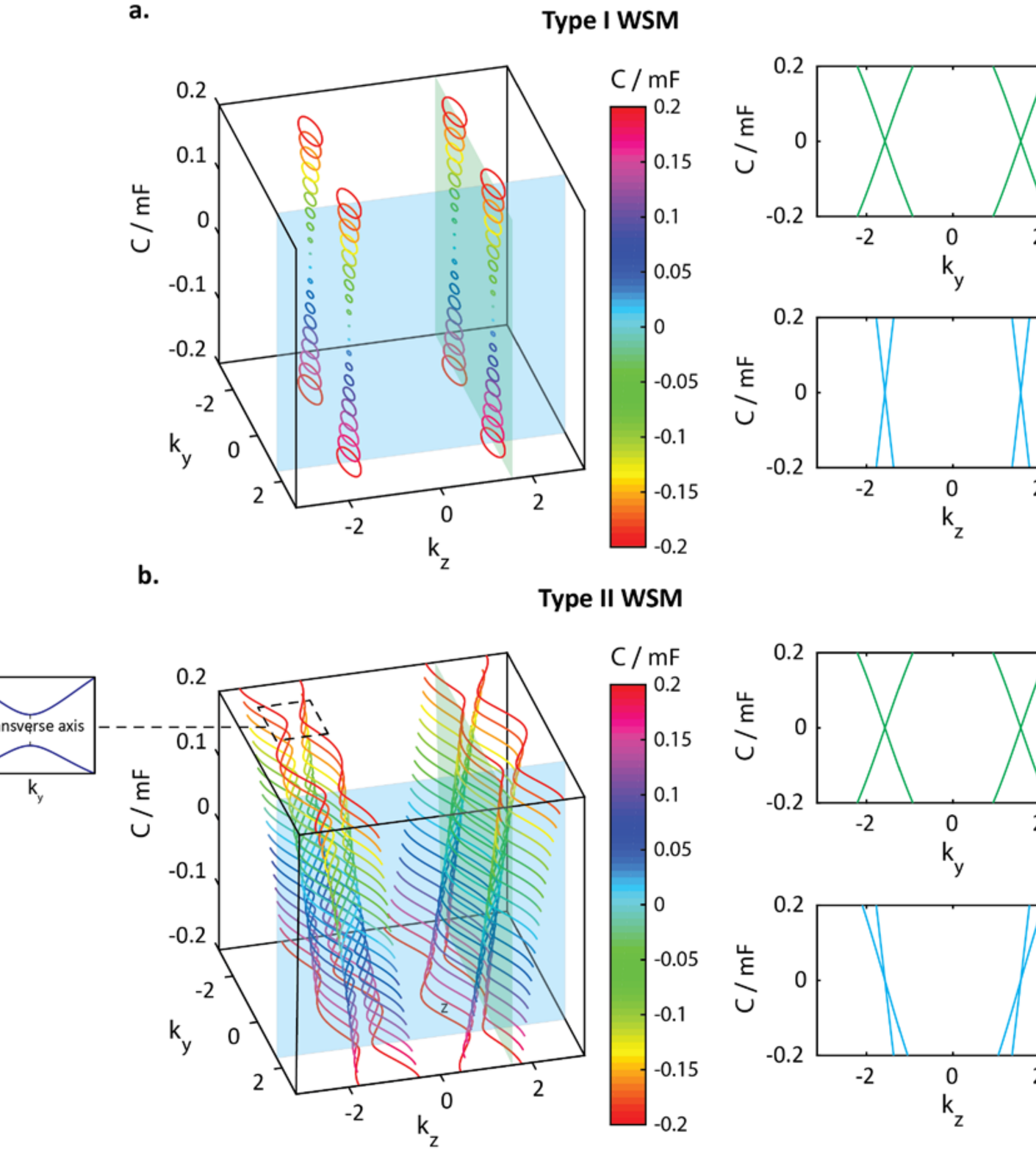}
\caption{ The equal-capacitance contours of a TE circuit of Fig. \ref{gFig2} consisting of  $C_1 = 0.716\ \mathrm{mF}$, $C_y = 0.167\ \mathrm{mF}$ and $C_{Az} = 0.5\ \mathrm{mF}$ at $k_x=0$ (main plot), $C$-dispersion relation with $k_y$ at $k_x=\pi,k_z=\pi/2$ (upper right plot) and $C$-dispersion relation with $k_z$ at $k_x=\pi,k_y=\pi/2$ (lower right plot) for a. a Type I WSM with $C_{Bz} = -0.5\ \mathrm{mF}$ and b. a Type II WSM with $C_{Bz} = 0.5\ \mathrm{mF}$. The $k_z=\pi/2$ and $k_y=\pi/2$ planes are indicated on the main plots. The inset of the main plot in panel b. shows a zoomed-in view of the hyperbolic ECC near one of the Weyl points.  } 
\label{gFig3}
\end{figure}

Fig. \ref{gFig3} shows the $C$ dispersion relations of an exemplary Type I WSM and Type II WSM modeled by the TE circuit of Fig. \ref{gFig2} with capacitive values $C_1 = 0.716\ \mathrm{mF}$, $C_y = 0.167\ \mathrm{mF}$ and $C_{Az} = 0.5\ \mathrm{mF}$, and $C_{Bz} = -0.5\ \mathrm{mF}$ for the Type I WSM and $C_{Bz} = 0.2\ \mathrm{mF}$ for the Type II WSM,  respectively.  In the respective circuits, the value of $C_0$ has been set so that the Weyl points occur at $C=0$.  (We chose $\mathrm{mF}$ as the units accompanying the numerical capacitance values used in our calculations because, for the reasonable value of $\omega = 1\ \mathrm{kHz}$, the inductance corresponding to a given negative numerical value of capacitance in $\mathrm{mF}$ has the same (positive) numerical value in $\mathrm{\mu H}$. Capacitors and inductors with capacitances and inductances on the order of $1\ \mathrm{mF}$ and $1\ \mathrm{\mu H}$,  respectively, would have physical dimensions of $\simeq 1\ \mathrm{cm}$, and are readily available. ) 

Fig. \ref{gFig3}a shows the untilted Type I WSM that results when $C_{Az}$ and $C_{Bz}$ have the same magnitudes, but opposite signs. The low-$C$ dispersion relation takes the familiar form of Dirac cones with ellipsoid cross sections centered around the four Weyl nodes. The gradients of the dispersion relation have opposite signs for opposite signs of $\delta k_y$ and $\delta k_z$ around each Weyl node. 

Fig. \ref{gFig3}b shows the Type II WSM that results when $C_{Az}$ and $C_{Bz}$ have the same signs. The equal capacitance contours (ECC, in analogue to the more familiar term `equal energy contours') are reflection-symmetric about the $k_y$ axis due to the  reflection symmetry about the $y$-axis in the TE circuit. The form of the ECCs in a Type II WSM stands in contrast to the more familiar ellipsoidal cross sections of the Type I WSM Dirac cones. In the vicinity of a Weyl node, the ECC of a Type II WSM takes the form of a hyperbola with its transverse axis  parallel to the tilt direction along the $z$ direction (see inset of Fig. \ref{gFig3}b). One branch of the hyperbola corresponds to hole states in the sense that if we write the Hamiltonian as $H = \vec{m}\cdot\vec{\sigma}$, then for an eigenstate on the hole branch, $|h\rangle$, we have $\langle h|\vec{\sigma}|h\rangle = -\hat{m}$. The other branch corresponds on particle-like states where an eigenstate on the branch, $|p\rangle$, yields $\langle p|\vec{\sigma}|p\rangle = \hat{m}$.  Unlike the ellipsoid ECCs of a Type I WSM, the hyperbola ECCs of the Type II WSM are not closed curves. The non-closure of hyperbolas and the periodicity of the first Brillouin zone together cause the ECCs to deviate from perfect hyperbolas far away from the projections of the Weyl nodes. In our TE circuit, the hyperbolas around Weyl nodes with the same sign of $k_z$ link up so that the ECCs take the form of lines running along the $k_y$ axis in the Brillouin zone. 

The side plots of panel b. illustrate an important distinction between a Type II WSM and a Type I WSM. Whereas the gradient of the dispersion relation around each Weyl node has opposite signs for opposite signs of $\delta k_y$ perpendicular to the tilt direction in both Type I and Type II WSMs, the gradients of the dispersion relations have the same signs for both signs of $\delta k_z$ parallel to the tilt direction in the Type II WSM. This characteristic property of a Type II WSM will lead to interesting consequences for its transport properties, as we shall see in the next section. 

\section{Energy Flux transport in Weyl heterojunction }
We now consider the transport across a heterojunction formed between a Type I and Type II WSMs shown in Fig. \ref{gFig3}. The  transmission of the energy flux is calculated for an incident mode from a semi-infinite long Type I source lead to a semi-infinite long Type II drain lead. This is the TE analogue of solving for the transmitted current for a given incident source mode in a TB heterojunction. In the Landauer-Buttiker picture conventionally used in calculating the conductances across heterojunctions in condensed matter systems, it is assumed that all states in the source lead are populated up to the source Fermi energy. In the limit of small applied source-drain bias, the calculation of the conductance requires the consideration of the contributions from all source states at the Fermi energy propagating from the source to the drain. However, in some cases, it is preferable to limit the conductance contributions to a restricted set of  source states at the  Fermi energy. For instance, in the case of the Datta-Das spin transistor \cite{APL56_665}, the ideal behaviour is approached by limiting the conductance contribution to normally incident source states. Thus, experimentally,  additional steps have to be taken to exclude the contributions of the unwanted source states. In contrast, the source lead in a TE system can be populated with only a single incident mode by appropriately setting the voltage biases at the source voltage nodes, as explained in the previous section. We therefore focus exclusively on the set of source modes with $k_x = \pi$ in the rest of the paper.  

\begin{figure}[ht!]
\centering
\includegraphics[scale=0.6]{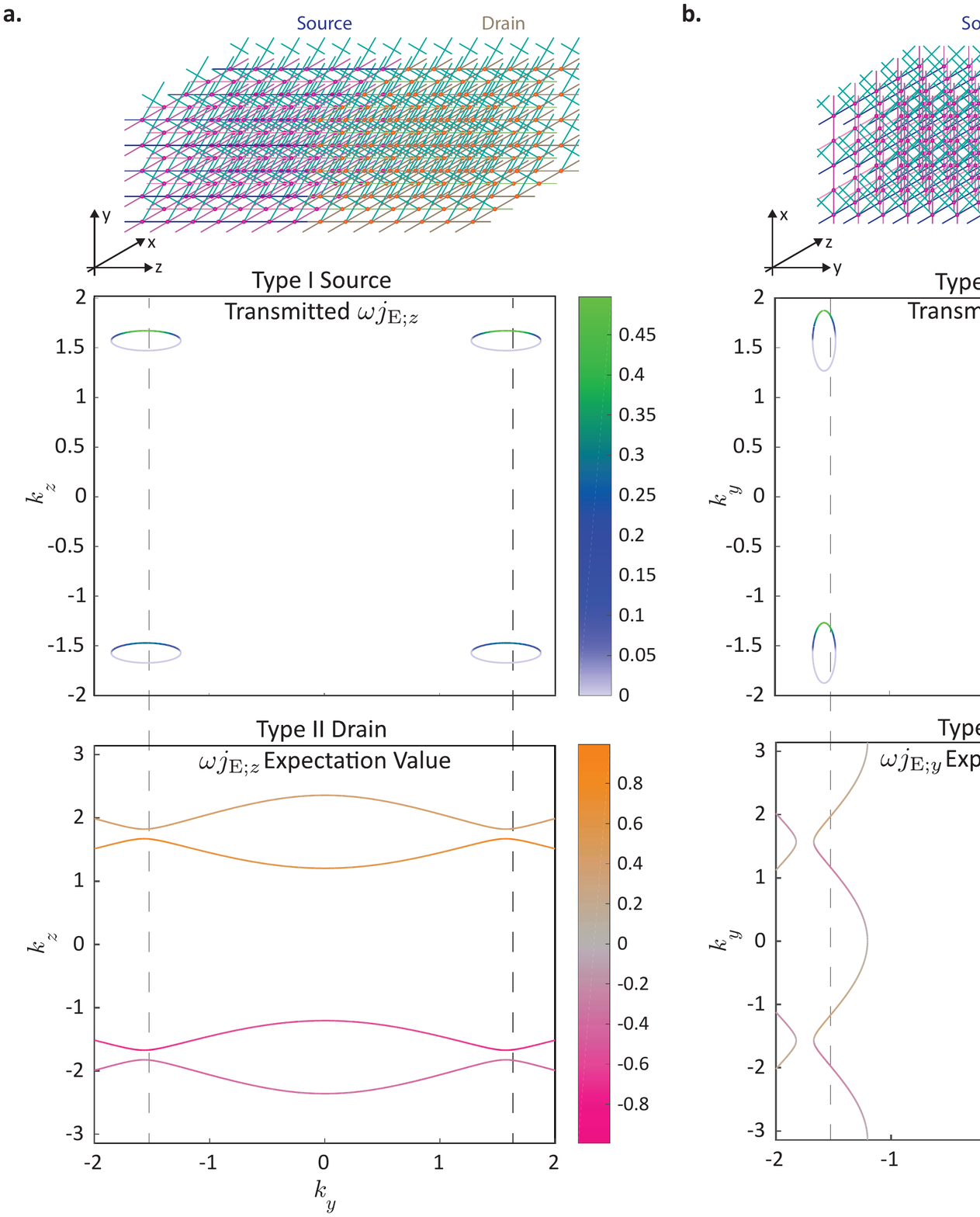}
\caption{The top row shows the three-dimensional schematic diagrams of the TE WSM circuits for energy flux propagation along a. the $z$ and  b. $y$ directions respectively. The dots represent the circuit nodes and the lines the capacitive couplings between the voltage nodes. The unconnected lines at the edges of the circuit indicate that the circuit extends to infinity beyond these lines. The second row shows the $k_x=\pi$ source ECCs at $C=0.1\ \mathrm{mF}$ and transmitted energy flux for each mode on the source ECC source incident on the source-drain interface between a Type I source lead and a Type II drain lead with the same parameters as in Fig. \ref{gFig3}. The third row shows the expectation values of the drain energy flux along the a. $z$, and b. $y$ directions. The vertical lines in panel a. and b. linking the second and third rows indicate points on the source and drain ECCs with the same conserved values of $k_y$ and $k_z$ momentum respectively.  (Note that the plot axes are swapped between panels a. and b. )} 
\label{gFig4}
\end{figure}	

We compare the two scenarios where the source-drain interface is parallel, and where the interface is perpendicular to the tilt direction along the $z$ direction. Fig. \ref{gFig4} shows the results of an exemplary calculation at $C=0.1\ \mathrm{mF}$ for a Type I source lead connected to a Type II drain lead with the same parameters as in Fig. \ref{gFig3}. The middle row of Fig. \ref{gFig4} shows the ECCs, and the transmitted energy flux at every point on the source ECCs for energy flux flowing in the $+z$ direction (a.) and in the $+y$ direction (b.). The transmitted energy flux was calculated by finding the linear superposition of source and drain eigenmodes which corresponds to flux incident on the source side, and satisfies the circuit KCL at every node in the circuit similar to what we did in Eqs. \ref{hetV0} to \ref{hetV1}. The transmitted energy flux is therefore equal to zero on one half of the source ECCs where the flux propagates away from the source-drain interface. The bottom row shows the drain ECCs and the energy flux along the $z$ direction (a.), and $y$ direction (b.) on each point of the ECCs. Denoting the eigenmode with eigenvalue $C$ at $\vec{k}$ as $\mathbf{v}(\vec{k})$, the energy flux along the $z$ and $y$ directions was calculated using $\mathbf{v}(\vec{k})^\dagger(\partial_{k_z/k_y} \mathbf{H})\mathbf{v}(\vec{k})$, which we refer to  this as the `expectation value' of the energy flux. We have oriented the axes of the plots so that the momentum perpendicular to the interface, which is conserved across the interface, lies on the horizontal axis of the respective  plots. The vertical lines spanning across the middle and bottom plots therefore link the source modes with the corresponding drain modes which they are transmitted into, subject to the conservation of transverse momentum. 

The transmission of the energy flux differs considerably between the scenarios where the transmission direction is parallel to, and when it is perpendicular to the tilt direction. In panel a. where the flux direction is parallel to the tilt direction, the transmitted energy flux for the source valleys with positive $k_z$ are considerably higher than the transmission for the source valleys with negative $k_z$. The valley asymmetry arises because of the distribution of the $z$-energy flux directions in the drain segment. At the given value of $C$, the drain segment hosts only modes which propagate in the $+z$ ($-z$) direction at positive (negative) $k_z$.  This is shown in the bottom plot in Fig. \ref{gFig4}a as well as the lower right inset of Fig. \ref{gFig3}b. The conservation of the transverse momentum $k_y$ restricts incident source modes at the source-drain interface to be transmitted into drain modes with the same value of $k_y$. This implies that source modes at the negative $k_z$ valley have to undergo inter-valley scattering into the positive $k_z$ drain valley in order to be transmitted into drain modes which propagate in the positive $z$ direction away from the interface. The large difference $\Delta k_z$ between the incident source mode and transmitted drain mode involved in the inter-valley scattering suppresses the transmission of source modes at the negative $k_z$ valleys relative to those at the positive $k_z$ valleys. For the positive $k_z$ source valley, the required $\Delta k_z$ in the \textit{intra}-valley scattering between the source and drain modes is much smaller.

The preferential transmission of energy flux for the positive $k_z$ valleys vanishes when the transmission direction (along $y$) is perpendicular to the tilt direction (along $z$). The transmitted energy flux in panel b. for propagation along the $+y$ direction has identical profiles in all four valleys, i.e. they are valley independent. Here, for a given conserved value of $k_z$, drain modes propagating in the $+y$ direction are present symmetrically around the vicinity of all four Weyl nodes. The source modes incident from all four valleys do not have to undergo inter-valley scattering in the drain segment in order to be transmitted in the forward $+y$ direction. 

A key distinction between a Type I WSM and a Type II WSM is that in the former (latter), there exists states in the vicinity of a Dirac point with both signs (one sign) of energy flux parallel to the tilt direction. This implies that at the transition from a Type I WSM to a Type II WSM, some of the states will have zero energy flux along the tilt direction. Let us consider the transition from a Type I to a Type II WSM. From Eq. \ref{T2Cond}, this transition occurs at $C_{Bz}=0$. In this special case, the $C$ dispersion relation of Eq. \ref{linHam}  takes the form of $C =  \eta_z (1  \pm 1) C_{Az} \delta k_z$ where the $\pm$ indicates the pseudospinor branch of the eigenvalue. $C$ does not disperse with $\delta k_z$ for the negative  pseudospinor branch. We term this transitionary regime between a Type I and Type II WSM  as Type III WSM. 

\begin{figure}[ht!]
\centering
\includegraphics[scale=0.5]{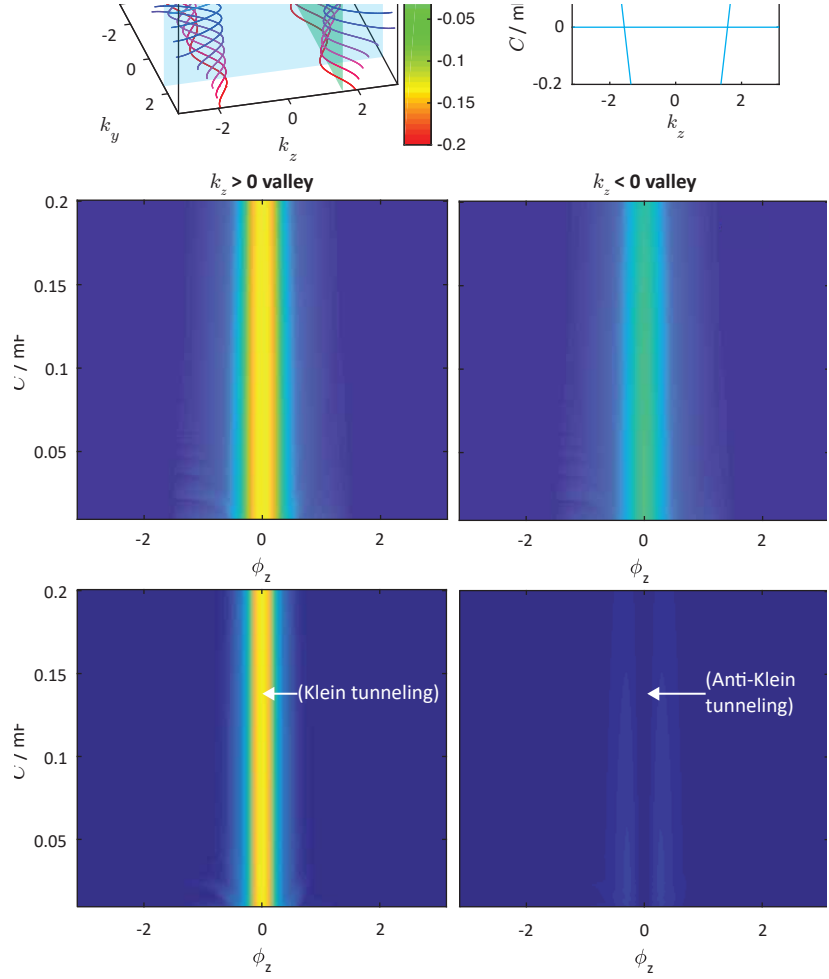} 
\caption{a. The ECCs for a $C_1=0.716\ \mathrm{mF}$, $C_y = 0.167\ \mathrm{mF}$, $C_{Az}= 0.5\ \mathrm{mF}$ and $C_{Bz} = 0\ \mathrm{mF}$ Type III WSM at $k_x = \pi$ (main plot), the $C$-dispersion relation with $k_y$ at $k_z = \pi/2$ (upper right plot) and $C$-dispersion relation with $k_z$ at $k_y=\pi/2$. The $k_z = \pi/2$ and $k_y=\pi/2$ planes are denoted in the main plot. b. and c. The variation of the transmitted energy flux with the common grounding capacitance $C$ and incidence angle $\phi_z$ from the positive $k_y$, positive $k_z$ (left column) valley and positive $k_y$, negative $k_z$ valley (right column) of a Type I WSM source to (b.) a Type II WSM drain lead with the parameters in panel a, and (c.) a Type III WSM drain lead with the parameters in Fig. \ref{gFig3}. }
\label{gFig5}
\end{figure}

Fig. \ref{gFig5}a shows the ECCs of a Type III WSM system (left), and the $k_y$ and $k_z$ dispersion relations cutting across the Weyl nodes (right). The lower right plot of panel a. shows that one of the branches of the cross section of the Dirac cone now lies flat parallel to the $k_z$ axis, in agreement with our prediction earlier from the linear expansion of $\delta\vec{k}$ around the Weyl points in  Eq. \ref{linHam}.  The  flatness of one of the eigenspinor branches means that for $k_x=\pi$ and $C \neq 0$, a line of constant $k_y$ cuts across the $C$ dispersion of  Type III WSM at two points instead of four points in the case of Type II WSM. Furthermore, the ECC profile in Fig. \ref{gFig5}a for a Type III WSM  consists of only two curves  across the $k_y$ range spanned by the Brillouin zone, instead of the four curves for a Type II WSM in Fig. \ref{gFig3}b. The absence of states from one of the eigenspinor branches leads to a qualitative difference for the transmission from a Type I WSM source lead to a Type III WSM drain lead, compared to the transmission from a Type I WSM source lead to a Type II drain lead. In what follows, we consider only the positive $k_y$ valleys because the reflection symmetry of the TE system along the $y$ direction results in the other pair of valleys with $k_y < 0$ having identical transmission profiles as their $y$ reflection partners. 

Panels b. and c. of Fig. \ref{gFig5} plot the transmission from the positive $k_y$, positive $k_z$ source valley (left column) and positive $k_y$, negative $k_z$ source valley (right column), from a Type I WSM source to a Type II WSM drain (panel b.), and to a Type III WSM drain (panel c.), as a function of the common capacitance $C$ and the incidence angle $\phi_z$.  An incidence angle of zero corresponds to normal incidence at the interface. For both Type II and Type III drain leads, the transmitted flux is higher for incident modes from the the positive $k_z$ valley than that from the negative $k_z$ valley. The suppressed transmission of the negative $k_z$ valley source modes is, in both cases, due to the inter-valley scattering into positive $k_z$ valley drain states required for transmission from the source to the drain. (The $k_z$ dispersion relation in the lower right plot in Fig. \ref{gFig5}a shows that in the vicinity of each Weyl point there are only modes with a single sign of $\partial_{k_z} C$, which signifies the energy flux along the $z$ direction.) However, a key difference between the transmission to a Type II drain as opposed to a Type III drain is that in the former, the transmission for the negative $k_z$ state is maximal at normal incidence at $k_z=0$ whereas in the latter, the transmission of the negative $k_z$ states at normal incidence is completely suppressed. This complete suppression of the transmission at normal incidence is, in a sense, the antithesis of the well-known Klein tunneling in Dirac fermion systems. We term this tunneling behaviour as ``anti-Klein tunneling". The emergence of  anti-Klein tunneling  is basically due to the fact that propagating states exist in only one eigenspinor branch around each Weyl point in a Type III WSM.

\begin{figure}[ht!]
\centering
\includegraphics[scale=0.6]{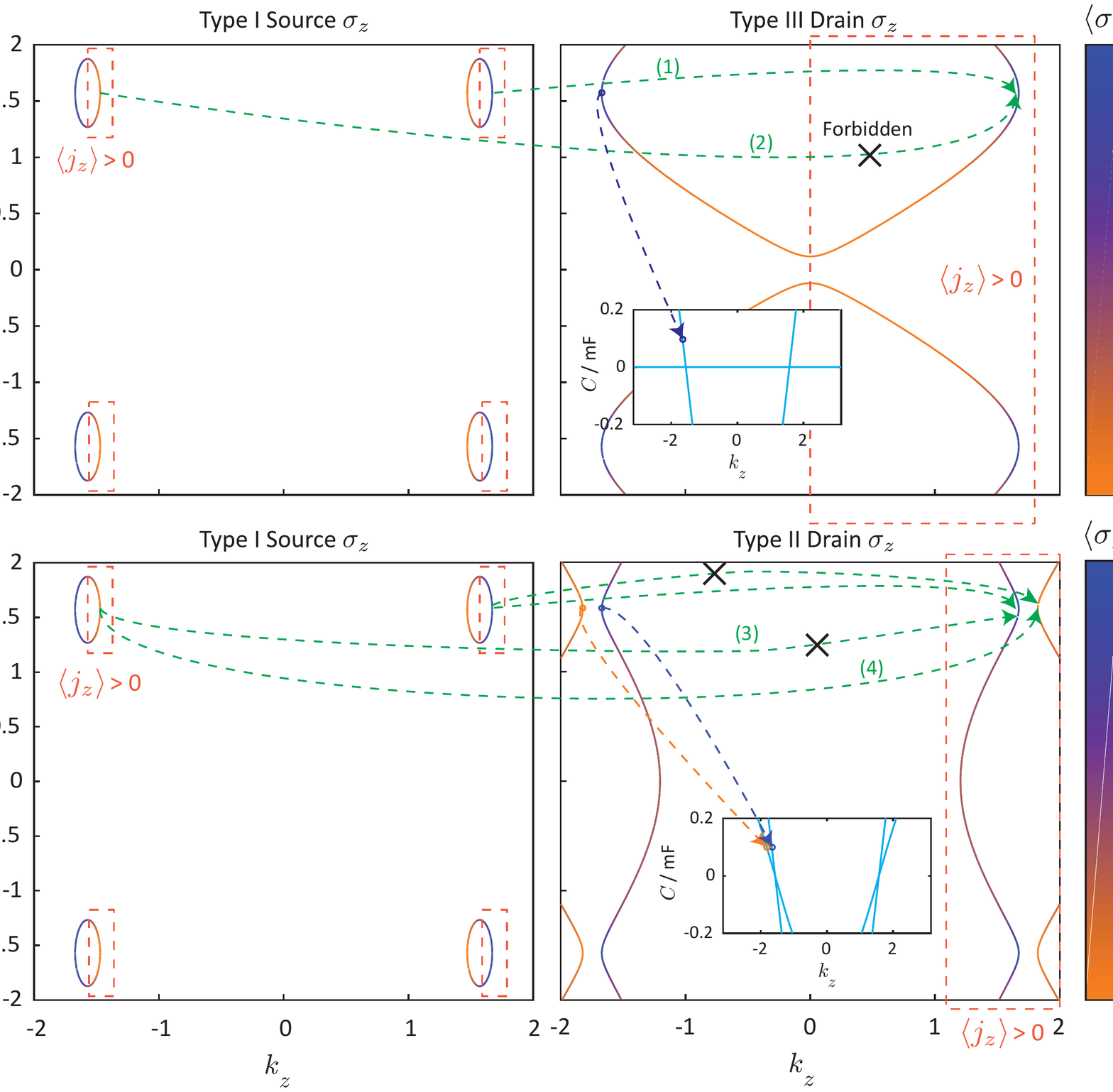} 
\caption{a. and b. The ECCs and $\langle \sigma_z \rangle$ at each mode on the ECCs in the source (left) and drain (right) at $C = 0.1$ for a. the Type III WSM drain and b. the Type II WSM drain of Fig. \ref{gFig5}. The red dotted boxes denote the modes with energy fluxes in the positive $z$ direction. The dotted arrows spanning across the source and drain ECCs at $k_y=\pi/2$ denote the drain states that the $\phi_z = 0$ source states can be transmitted into. The insets in the right columns relate the ECC curves to bands on the $C-k_z$ dispersion relation. The labels (1) to (4) denote various allowed and forbidden transmission processes from the positive $k_y$, $\delta k_y=\delta k_x = 0$ source modes in the two valleys to the forward-propagating drain modes. }
\label{gFig6}
\end{figure}	

To further explain the existence of the anti-Klein tunneling behaviour, let us consider Fig. \ref{gFig6} which shows the ECCs in the source (left column) and drain (right column) for a Type III WSM drain lead (panel a.), and a Type II WSM drain lead (panel b.), and the expectation value of the pseudospin $\sigma_z$ at each point of the ECC. The modes which propagate in the positive $z$ direction from the source to the drain ( the right halves of the elliptical ECCs in the Type I source, and the positive $k_z$ curves in the Type II and Type III drains) are demarcated in dotted red boxes in the figure. As noted earlier in the discussion on Fig. \ref{gFig5}, the Type III ECCs at any non-zero value of $C$ consist of only two curves running along the $k_y$ axis, both of which belong to the $+\sigma_z$ eigenspinor  branch. The normal incidence of the source mode from each valley at $\phi_z=0$ corresponds to the conserved values of $\delta k_x=\delta k_y=0$ around each Weyl point in both the source and drain segments in the linear $\delta\vec{k}$ expansion Eq. \ref{linHam}.  At normal incidence and $C_{Bz}=0$, Eq. \ref{linHam} reduces to $H = -C_{Az} \eta_z \delta k_z (\mathbf{I}_\sigma + \sigma_z)$. Since the given value of $C > 0$, $-2C_{Az}\eta_z\delta k_z > 0$. The sign of the pseudospin $z$ expectation value of the $+\sigma_z$ eigenstate follows that of $-C_{Az} \eta_z \delta k_z$. The pseudospin about the Weyl points at $\delta k_x=\delta k_y = 0$ would thus lie along the  $+z$ direction for all four drain valleys in Type III WSM, as shown in the right column of panel a.  This matches the $\sigma_z$ direction  of the normally incident source states from the positive $k_z$ valleys (left plot of panel a), which thus get transmitted perfectly (transition (1) in Fig. \ref{gFig6}). However, the normally incident source states from the negative $k_z$ valleys have pseudospin $\sigma_z$ along $-z$, i.e., in the opposite direction to that of the drain states. In other words, the source and drain states at normal incidence are orthogonal to one other as far as their pseudospin state is concerned. This leads to a total suppression of the transmission of normally incident states, i.e., ``anti-Klein tunneling" (the forbidden transition (2)). In conventional Klein tunneling with coincident source and drain Dirac points, there is perfect transmission because the pseudospin directions of the source and drain states point in exactly the same direction regardless of the potential difference between the source and drain. In contrast, in our TE (Type I-Type III) heterojunction system, the pseudospin direction of the normally incident source states at the negative $k_z$ point in exactly opposite direction to that of the corresponding drain states, thus resulting in zero transmission. 

In contrast to the Type III WSM drain segment in panel a. of Fig. \ref{gFig6}, the Type II drain segment in panel b. has states from both eigenspinor ($\sigma_z=\pm 1$) branches present at all four Weyl points. The transmission of normally incident modes from the negative $k_z$ valley to the forward propagating drain modes in the positive $\sigma_z$ branch (the forbidden transition (3).) is prohibited due to their respective pseudospins pointing in opposite directions. However, it is possible for the source modes to be transmitted into the  drain modes in the $\sigma_z = -1$ branch. Indeed at normal incidence, the pseudospin directions of the source modes and the corresponding drain modes in the $ -\sigma_z$ branch point in exactly the same direction, resulting in the  valley transmission peaking at $\phi_z=0$ for the $-k_z$ valley (transition (4)). However, since inter-valley scattering is involved in the transmission, the transmission probability $T$ is somewhat suppressed, in contrast to the perfect transmission in conventional Klein tunneling.

\section{Conclusion}
In this work, we first established the analogy between a TE circuit and a TB system. We saw that the common grounding capacitance and the energy flux in the former are analogous to the eigenenergy and the probability flux in the latter. We also described how multi-dimensional heterojunctions between semi-infinite leads may be modeled using a finite number of nodes. 

Exploiting the analogy between LC TE circuits and TB Hamiltonians, we proposed the realization of various WSM phases to construct a WSM heterojunction between a Type I WSM source and a Type II/III WSM drain. A TE circuit offers additional flexibility and tunability over a condensed matter system where in the former, a source lead can be populated with a specific mode by adjusting the voltage bias appropriately. In contrast, in a condensed matter system, the conductance is contributed by all source states at the source Fermi energy, and it is difficult to restrict the conductance to a desired subset of source modes. We saw that the transport properties differ significantly for the cases when the transport direction is parallel to, and when it is perpendicular to the $k$-space tilt direction.  By investigating the relative orientation between tilt and flux direction, we observed that large inter-valley scattering suppresses the transmission for one pair of source valleys relative to the other when the tilt and flux direction are parallel to each other. However, all valleys contribute equally in transmission when the tilt direction is perpendicular to the propagation direction. Moreover, by adjusting the coupling capacitance, we achieved an exotic Type III WSM drain lead where one of the eigenspinor branches shows zero group velocity along the tilt direction. The flatness of Type III WSM  band dispersion results in anti-Klein tunneling where the transmission for the negative $k_z$ valleys at normal incidence is totally suppressed. This is due to the pseudospin directions of the source and drain modes pointing in exactly opposite directions, leading to the orthogonality of the source and drain pseudospin states. 

Our proposed model can be easily implemented experimentally with conventional electrical components. Realizing various WSM phases in a single real material is difficult to achieve in practice. We thus propose the TE circuit network analog of a WSM heterojunction as a platform to study the transport behavior of WSM materials. 

\section{Acknowledgments}
This work is supported by the Singapore National Research Foundation (NRF), Prime Minister’s Office, under its Competitive Research Programme (NRF CRP12-2013-01, NUS Grant No. R-263-000-B30-281), as well as the MOE Tier-I FRC grant (NUS Grant No. R‐263‐000‐D66‐114).

\section*{Appendix}
We mentioned earlier that it is not possible to uniquely set the voltage profile in an infinite one-dimensional chain by attaching a finite number of voltage supplies. We explain why this is the case here.

\begin{figure}[ht!]
\centering
\includegraphics[scale=0.6]{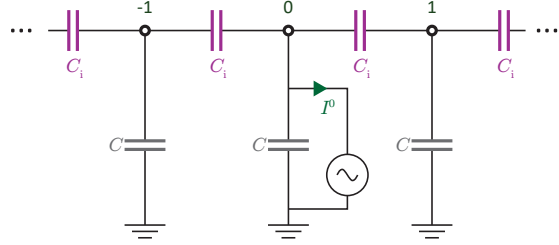} 
\caption{ An infinite chain consisting of voltage nodes coupled to their left and right neighbours via a coupling capacitance $C_{\mathrm{i}}$ and to the ground via the common grounding capacitance $C$. A voltage supply is attached to the $n=0$ node with the current flowing through the supply denoted as $I^0$.    }
\label{gFigApp1}
\end{figure}	

Fig. \ref{gFigApp1} shows the same infinite chain as Fig. \ref{gFig1}b except that we now attach a voltage supply at the $n=0$ node. Due to the periodicity of the chain away form $n=0$, we can write the voltage profile at $n<0$ to the left of the voltage supply as 
\begin{equation}
	V_n = \alpha_+ \exp(i \kappa n) + \alpha_- \exp(-i \kappa n), \ n < 0 \label{appE1}
\end{equation}
and the profile at $n >0$ to the right of the voltage supply as 
\begin{equation}
	V_n = \beta_+ \exp(i \kappa n) + \beta_- \exp(-i \kappa n), \ n > 0 \label{appE2} 
\end{equation}
where $\alpha_\pm$ and $\beta_\pm$ are coefficients for the $\exp(\pm i \kappa n)$ modes to the left and right of the voltage supply respectively, and $\kappa \equiv \arccos( \frac{C}{2C_\mathrm{i}} + 1 )$. In general, $\alpha_\pm \neq \beta_\pm$. After setting the voltage at node $n=0$ by $V_0$, there would be a total of five unknowns consisting of $\alpha_\pm$, $\beta_\pm$ and $I^0$, the current flowing through the voltage supply. It turns out that whereas the KCLs at nodes -1, 0 and 1 yield meaningful equations, the KCLs at nodes $n$, $|n| > 1$ do not. To see this, consider the KCL at $n=3$. 
\begin{eqnarray}
	&& C V_3 = C_{\mathrm{i}} (V_2 + V_4 - 2 V_3) \label{appE2a} \\
	&\Rightarrow& (\beta_+ \exp(3i\kappa) + \beta_-\exp(-3i\kappa)) ( C_\mathrm{i} (2\cos(\kappa-1) - C) = 0 \label{appE2b}
\end{eqnarray}
where in Eq. \ref{appE2b}, we have made use of Eq. \ref{appE2} to express $V_{2,3,4}$ in terms of $\beta_\pm$. Eq. \ref{appE2b} reduces to the trivial equation $0=0$ because by definition $\kappa$ satisfies $ ( C_\mathrm{i} (2\cos(\kappa-1) - C)  = 0$. 
Thus applying KCL at all the nodes would yield only three equations for five unknowns. In other words, setting the voltage bias to be $V_0$ at node $n=0$ does not uniquely specify the voltage profile in the infinite chain. For example, if we have substituted $\beta_\pm = \alpha_\pm$ into Eq. \ref{appE1} and \ref{appE2} and then solved for the resulting system of equations comprising the KCLs at nodes -1, 0 and 1 in terms of $\alpha_\pm$, we would have obtained the solutions $I^0=0, \alpha_\pm = V_0/2$. However, a different solution results if we adopt a different sequence: We can leave the KCLs at node $n=\pm 3$ in terms of $\kappa$  (like Eq. \ref{appE2b} for $n=3$) and solve the system of equations consisting of the KCLs at nodes -2 to 2 so as to obtain solutions of the unknowns explicitly in terms of $\kappa$. If we now substitute $\kappa = \arccos( \frac{C}{2C_\mathrm{i}} + 1 )$ into the solutions, we would have ended up with a different voltage profile with a finite current flowing through the voltage supply. 

Adding in more voltage supplies to the chain does not fix the problem. Each additional voltage supply comes with the additional unknown of the current flowing through the voltage supply, so there will always be more unknowns than equations.

\end{document}